\documentclass[12pt,draft]{article}

\usepackage{amsthm,amssymb,amsmath,amscd,amstext,comment,latexsym,mathrsfs}
\usepackage{mathrsfs,mathptmx,tikz,xypic}

\newtheorem{definition}{Definition}[section]
\newtheorem{theorem}{Theorem}[section]
\newtheorem{lemma}{Lemma}[section]
\newtheorem*{remark}{{\it Remark}}

\newcommand{\nc}{\newcommand}

\nc{\C}{{\mathbb C}}
\nc{\R}{{\mathbb R}}
\nc{\HH}{{\mathbb H}}
\nc{\Z}{{\mathbb Z}}
\nc{\N}{{\mathbb N}}

\nc{\dd}{{\rm d}}
\nc{\ii}{{\bf i}}
\nc{\ch}{{\mathscr H}}
\nc{\ci}{{\mathscr I}}
\nc{\cj}{{\mathscr J}}

\nc{\fb}{{\mathfrak B}}
\nc{\fii}{{\mathfrak I}}
\nc{\fj}{{\mathfrak J}}
\nc{\fr}{{\mathfrak R}}
\nc{\fm}{{\mathfrak M}}

\nc{\sm}{{\mathsf M}{\mathsf a}{\mathsf n}^4}

\DeclareMathOperator{\ed}{End}

\nc{\stack}[2]{{\begin{array}{c}
\scriptstyle #1 \\ \scriptstyle #2 \end{array}} }

\begin{document}

\title{The universal von Neumann algebra of smooth four-manifolds revisited}

\author{G\'abor Etesi\\
\small{{\it Department of Algebra and Geometry, Institute of Mathematics,}}\\
\small{{\it Budapest University of Technology and Economics,}}\\
\small{{\it M\H uegyetem rkp. 3., H-1111 Budapest, Hungary}}
\footnote{E-mail: {\tt etesi@math.bme.hu}}}

\maketitle

\pagestyle{myheadings}
\markright{The universal von Neumann algebra of smooth four-manifolds revisited}

\thispagestyle{empty}

\begin{abstract} 
Making use of its smooth structure only, out of a 
connected oriented smooth $4$-manifold a von Neumann algebra is 
constructed. It is geometric in the sense that is generated by local 
operators and as a special four dimensional phenomenon it contains all 
algebraic (i.e., formal or coming from a metric) curvature tensors of the 
underlying $4$-manifold. The von Neumann algebra itself is a hyperfinite 
factor of ${\rm II}_1$-type hence is unique up to abstract isomorphisms of 
von Neumann algebras. Over a fixed $4$-manifold this universal von Neumann 
algebra admits a particular representation on a Hilbert space such that its 
unitary equivalence class is preserved by orientation-preserving 
diffeomorphisms consequently the Murray--von Neumann coupling constant of 
this representation is well-defined and gives rise to a new and computable 
real-valued smooth $4$-manifold invariant. Its link with Jones' subfactor 
theory is noticed as well as computations in the simply connected closed 
case are carried out.

Application to the cosmological constant problem is also discussed. 
Namely, the aforementioned mathematical construction allows to 
reformulate the classical vacuum Einstein equation with cosmological 
constant over a $4$-manifold as an operator equation over its tracial 
universal von Neumann algebra such that the trace of a solution is naturally 
identified with the cosmological constant. This framework permits to use 
the observed magnitude of the cosmological constant to estimate by 
topological means the number of primordial black holes about the Planck era. 
This number turns out to be negligable which is in agreement 
with known density estimates based on the Press--Schechter mechanism.

\end{abstract}

\centerline{AMS Classification: Primary: 46L10, 83C45, Secondary: 57R55}

\centerline{Keywords: {\it Smooth $4$-manifold; Hyperfinite ${\rm II}_1$
factor; Quantum gravity; Cosmological constant}}

%%%%%%%%%%%%%%%%%%%%%%%%%%%%%%%%%%%%%%%%%

\section{Introduction and summary}
\label{one}

%%%%%%%%%%%%%%%%%%%%%%%%%%%%%%%%%%%

The 20-$21^{\rm st}$ century has been witness to a great expansion of
mathematics and physics bringing a genuinely two-sided 
interaction between them. The 1980-90's culmination of discoveries
in low dimensional differential topology driven by Yang--Mills
theory of particle physics has dramatically changed our understanding of
four dimensional spaces: nowadays we know that the interplay between
topology and smoothness is unexpectedly complicated precisely in
four dimensions leading to the existence of a superabundance of smooth
four dimensional manifolds. While traditional invariants
of differential topology loose power in three and four dimensions, the new
quantum invariants provided by various Yang--Mills theories work
exactly in these dimensions allowing an at least partial enumeration
of manifolds. It is perhaps not just an accidence that
quantum invariants are applicable precisely in three and four dimensions,
equal to the phenomenological dimensions of physical space and space-time.

It is interesting that unlike Yang--Mills theories, classical general
relativity---despite its powerful physical content, too---has not
contributed to our understanding of four dimensionality yet. This might
follow from the fact that general relativity, unlike Yang--Mills theories
with their self-duality phenomena, permits formulations in every
dimensions greater than four exhibiting essentially the same properties.
This is certainly true when general relativity is considered in its usual
fully classical differential-geometric school uniform however four dimensionality
enters the game here as well if one tries to link differential geometry
with non-commutativity \cite{con}. 

The purpose of these notes is therefore twofold. On the {\it mathematical side} 
they intend to replace in a substantionally new way the overcomplicated, 
superfluous and, to confess, at many technical steps erroneous construction 
exhibited in our earlier work \cite[Lemmata 2.1, 2.2 and 2.3]{ete3}. 
The aim of that construction was to extract from a connected oriented smooth 
$4$-manifold $M$ a ${\rm II}_1$ hypefinite factor von Neumann 
algebra $\fr$. The independent new construction here has exactly the same 
purpose however, in sharp contrast to our earlier struggles \cite{ete2, ete3}, 
it rests on a standard Clifford algebra construction only hence 
is both conceptionally and technically very simple; thereby reducing the 
status of the machinery $M\Rightarrow\fr$ to a sort of 
so-far-unnoticed triviality. In addition, as a further novelty, 
a connection with the associated smooth $4$-manifold invariant 
$\gamma$ introduced and studied in \cite[Lemmata 2.4 and 2.5]{ete3} with Jones' 
subfactor theory is observed and new computations with this invariant in the 
simply connected closed case are exhibited. 

Our mathematical results are summarized as follows:

\begin{theorem} Let $M$ be a connected oriented smooth $4$-manifold.
Making use of its smooth structure only, a von Neumann algebra $\fr$
can be constructed which is geometric in the sense that it is generated by 
local operators including all bounded complexified algebraic 
(i.e., formal or stemming from a metric) curvature tensors of $M$ and 
$\fr$ itself is a hyperfinite factor of type ${\rm II}_1$ hence is unique 
up to abstract isomorphism of von Neumann algebras.

Moreover $\fr$ admits a representation on a certain 
separable Hilbert space over $M$ such that the unitary equivalence class 
of this representation is invariant under orientation-preserving 
diffeomorphisms of $M$. Consequently the Murray--von Neumann coupling constant 
of this representation gives rise to a smooth invariant 
$\gamma (M)\in [0,1)$. More precisely it satisfies 
\[\gamma(M)=1-\frac{1}{x}\] 
where $x\in\{4\cos^2\big(\frac{\pi}{n}\big)\:\vert\:n\geqq 3\}\cup[4,+\infty)$, 
the set of Jones' subfactor indices. 
\label{nagytetel1}
\end{theorem}
\noindent This invariant has the following basic properties which are helpful 
in computations:

\begin{theorem} The invariant behaves like $\gamma(\overline{M})=\gamma (M)$ 
under reversing orientation, $\gamma (M\setminus Y)=
\gamma (M)$ under excision of homologically trivial submanifolds and 
\[\gamma (M\#N)=\frac{\gamma (M)+\gamma (N)}{1+\gamma (M)\gamma (N)}\] 
under connected sum. 
\label{nagytetel2}
\end{theorem}

\noindent The invariant in the closed simply connected case can be 
characterized with the following properties. These show that 
unfortunately $\gamma$ is not sensitive enough yet i.e., its construction 
demands further improvements in order to effectively distinguish smooth 
structures:

\begin{theorem} If $M', M''$ are connected, closed, simply
connected, smooth $4$-manifolds which are homeomorphic then 
$\gamma (M')=\gamma (M'')$. 

More precisely if $M$ is a closed, simply connected, smooth $4$-manifold then 
\[\gamma (M)=\frac{17^{b_2(M)}-1}{17^{b_2(M)}+1}\]
holds.
\label{nagytetel3}
\end{theorem}

\noindent Thus for example $\gamma (S^4)=0,\gamma (\C P^2)=\frac{8}{9}$ 
and $\gamma (\C P^1\times\C P^1)=\frac{144}{145}$. We also find for instance 
for the $K3$ surface having $b_2(K3)=22$ that already  
%$\gamma (K3)=R_{22}\left(\frac{8}{9}\right)=
%\frac{2^4\cdot3^2\cdot23\cdot947\cdot87415373\cdot2141993519227}
%{5\cdot29\cdot89\cdot25741\cdot256152733\cdot6901823633}$ 
%which is equal to 
%$1-\frac{1}{5\cdot29\cdot89\cdot25741\cdot256152733\cdot6901823633)}$ 
$1-\gamma (K3)\approx1.70\times10^{-27}$ and the general asymptotics of 
$\gamma$ in the simply connected case is 
$0<1-\gamma (M)\approx{\rm e}^{-{\rm const.}\:b_2(M)}$. This indicates that 
this invariant maps four dimensional smooth structures into $[0,1)$ in a 
logarithmic way in some sense.

Let us add some comments here. The set of 
Jones indices splits into a discrete and a continuous part. Subfactors 
belonging to the discrete part 
$\{4\cos^2\big(\frac{\pi}{n}\big)\:\vert\:n\geqq 3\}$ have been
completely classified \cite{pop} and in turn they follow an $ADE$
pattern (with the odditiy that no subfactors
corresponding to $D_{2k+1}$ and $E_7$ exist) \cite{ocn}. The set of subfactors 
belonging to the continuous portion $[4,+\infty)$ is however very wild and only 
partial results are known mainly for the subinterval $[4,5]\subset [4,+\infty)$ 
or a bit more (cf. e.g. \cite{jon-mor-sny} for a survey and results). The 
connected sum formula exhibited here in Theorem \ref{nagytetel2} shows that 
all non-prime $4$-manifolds (i.e., which admit non-trivial connected sum 
decompositions) belong via their invariant to the continuous regime i.e., 
$\frac{1}{1-\gamma (M)}\in [4,+\infty)$ furthermore we also find 
by Theorem \ref{nagytetel3} that this is also true for the prime manifold 
complex projective plane 
because $\frac{1}{1-\gamma (\C P^2)}=9$. This observation strongly hints 
that smooth $4$-manifolds might provide a rich reservoir of 
subfactors in the wild index range moreover poses the question whether 
or not smooth $4$-manifolds corresponding to the tame (i.e., discrete) 
range exist (e.g. $\gamma (S^4)=0$ corresponds to the Jones index 
$4\cos^2\big(\frac{\pi}{3}\big)=1$ hence the trivial case $n=3$ in the 
discrete part is realized by some $4$-manifolds). 

The other purpose of these notes, on the {\it physical side}, is to 
provide further clarifications to the material collected in 
\cite{ete2,ete3} regarding how to interpret the merely mathematical 
considerations above within a physical theory as naturally as possible. 
The outstanding problem of contemporary theoretical physics is how to 
unify the obviously successful and mathematically consistent {\it theory 
of general relativity} with the obviously successful but yet 
mathematically problematic {\it relativistic quantum field theory}. It 
has been generally believed that these two fundamental pillars of modern 
theoretical physics are in a clash not only by the different 
mathematical tools they use but even at a deep conceptional level: 
classical notions of general relativity such as the space-time event, 
the light cone or the event horizon of a black hole are too ``sharp'' 
objects from a quantum theoretic viewpoint meanwhile relativistic 
quantum field theory is not background independent from the aspect of 
general relativity.

Our interpretational efforts naturally fit into this context: they rest 
on the observation made explicit already in Theorem \ref{nagytetel1} 
that the constructed operator algebra $\fr$ is unique i.e. independent of the
particular manifold we have started with and, as a curiosity of four 
dimensions, it contains among its members algebraic curvature tensors 
over $M$. Since curvature plays a 
central role in classical general relativity it is quite natural to 
place this interpretation within the broad realm of quantum gravity. 
More concretely first by recalling the basic dictionary from 
\cite[Section 3]{ete2} we exhibit a straightforward generalization of 
the classical vacuum Einstein equation with cosmological constant over a 
$4$-manifold $M$ to an operator equation over its von Neumann algebra 
$\fr$ (see Definition \ref{kvantumeinstein} below). One interesting 
feature of this generalization is that the trace of those operators 
which satisfy this ``quantum vacuum Einstein equation'' can be 
identified with the cosmological constant. Second, we use the previous 
mathematical results to find solutions to this equation whose traces 
(hence the cosmological constants they give rise to) are equal to the 
smooth invariant $1-\gamma$ of Theorem \ref{nagytetel1} hence fall into 
$(0,1]\subset\R$. Consequently they are always {\it strictly positive} 
but {\it small} numbers (cf. \cite{ass-kro} too). Third, the observed 
small positive value of the cosmological constant \cite{rie-etal} allows 
in this framework to obtain an estimate on the number of primordial 
black holes in the very early Universe; the magnitude of this number 
turns out to be $\sim 10^2$ hence their presence is negligable which is 
in accord with the current conviction in the cosmologist and perhaps 
particle physicist community \cite{car-kuh}. Other recent works treating 
hyperfinite ${\rm II}$ factors from a gravitational perspective are 
\cite{ass-kro-wil, cha-lon-pen-wit}.

The present paper is heavily based on our earlier attempts \cite{ete2, 
ete3} but with substantial clarifications and new extensions. It is 
organized as follows. Section \ref{two} is a self-contained presentation 
of all the mathematical constructions involved in Theorems 
\ref{nagytetel1}, \ref{nagytetel2} and \ref{nagytetel3} together with 
their proofs. The language of this part is therefore rather 
mathematical. Then Section \ref{three} contains the physics by offering 
a quantum generalization of the classical vacuum Einstein equation (see 
Definition \ref{kvantumeinstein}) with an application to the 
cosmological constant problem. The language of this part is therefore 
rather physical. The physicist-minded reader may skip Section \ref{two} 
and read Section \ref{three} first hopefully without problem.

%%%%%%%%%%%%%%%%%%%%%%%%%%%%%%%%%%%%%%%%%

\section{Mathematical constructions}
\label{two}

%%%%%%%%%%%%%%%%%%%%%%%%%%%%%%%%%%%

In this section we shall first exhibit a simple self-contained two-step 
construction of a von Neumann algebra attached to any oriented smooth 
$4$-manifold. The structure of this algebra will be also explored in some 
detail. This is then will be followed by introducing a new smooth $4$-manifold 
invariant whose simplest properties will be examined too. 

{\it Construction of an algebra.} Take the isomorphism class of a connected 
oriented smooth $4$-manifold (without boundary) and from now on let $M$ be a 
once and for all fixed representative in it carrying the action of its own 
orientation-preserving group of diffeomorphisms ${\rm Diff}^+(M)$. Among all 
tensor bundles $T^{(p,q)}M$ over $M$ the $2^{\rm nd}$ exterior power 
$\wedge^2T^*M\subset T^{(0,2)}M$ is the only one which can be endowed with a 
pairing in a natural way i.e., with a pairing extracted from the smooth 
structure (and the orientation) of $M$ alone. Indeed, consider its associated 
vector space $\Omega^2_c(M):=C^\infty _c(M;\wedge^2T^*M)$ of compactly 
supported smooth $2$-forms on $M$. Define a pairing $\langle\:\cdot\:, 
\:\cdot\:\rangle_{L^2(M)}:\Omega^2_c(M)\times \Omega^2_c(M) 
\rightarrow\R$ via integration: 
\begin{equation}
\langle\alpha,\beta\rangle_{L^2(M)}:=
\int\limits_M\alpha\wedge\beta
\label{integralas}
\end{equation}
and observe that this pairing is non-degenerate however is 
{\it indefinite} in general thus can be regarded as an indefinite 
scalar product on $\Omega^2_c(M)$. It therefore induces an indefinite real 
quadratic form $Q$ on $\Omega^2_c(M)$ given by 
$Q(\alpha):=\langle\alpha,\alpha\rangle_{L^2(M)}$. Let $C(M)$ denote 
the complexification of the infinite dimensional real Clifford algebra 
associated with $(\Omega^2_c(M),Q)$. Because Clifford algebras are usually 
constructed out of definite quadratic forms, we summarize 
this construction \cite[Section I.\S 3]{mic-law} to make sure that 
the resulting object $C(M)$ is well-defined i.e. is not sensitive for the 
indefiniteness of (\ref{integralas}). To begin with, let 
$V_m\subset\Omega^2_c(M)$ be an $m$ dimensional 
real subspace and assume that $Q_{r,s}:=Q\vert_{V_m}$ has signature $(r,s)$ on 
$V_m$ that is, the maximal positive definite subspace of $V_m$ with respect to 
$Q_{r,s}$ has dimension $r$ while the dimension of the maximal negative 
definite subspace is $s$ such that $r+s=m$ by the non-degeneracy of 
$Q_{r,s}$. Then out of the input data $(V_m, Q_{r,s})$ one constructs in the 
standard way a finite dimensional real Clifford algebra $C_{r,s}(M)$ with unit 
$1\in C_{r,s}(M)$ and an embedding $V_m\subset C_{r,s}(M)$ with the property 
$\alpha^2=Q_{r,s}(\alpha)1$ for every element $\alpha\in V_m$. This real algebra 
depends on the signature $(r,s)$ however fortunately its complexification 
$C_m(M):=C_{r,s}(M)\otimes\C$ is already independent of it. In fact, 
if $\fm_k(\C)$ denotes the algebra of $k\times k$ complex matrices, then it is 
well-known \cite[Section I.\S 3]{mic-law} that $C_0(M)\cong\fm_1(\C)$ while 
$C_1(M)\cong\fm_1(\C)\oplus\fm_1(\C)$ and the higher dimensional cases follow 
from the complex periodicity $C_{m+2}(M)\cong C_m(M)\otimes\fm_2(\C)$. 
Consequently depending on the parity $C_m(M)$ is isomorphic to either 
$\fm_{2^{\frac{m}{2}}}(\C)$ or 
$\fm_{2^{\frac{m-1}{2}}}(\C)\oplus\fm_{2^{\frac{m-1}{2}}}(\C)$. These imply 
that $2$-step-chains of successive embeddings of real subspaces 
$V_m\subset V_{m+1}\subset V_{m+2}\subset\Omega^2_c(M)$ starting with 
$V_0=\{0\}$ and given by iterating $\omega\mapsto\binom{\omega}{0}$ provide us 
with injective algebra homomorphisms 
$\fm_{2^{\frac{m}{2}}}(\C)\hookrightarrow \fm_{2^{\frac{m}{2}+1}}(\C)$ having 
the shape $A\mapsto\left(\begin{smallmatrix} A & 0\\
                             0 & A
               \end{smallmatrix}\right)$. 
Therefore $C(M)$ is isomorphic to the injective limit of this directed system, 
that is there exists a linear-algebraic isomorphism 
\begin{equation}
C(M)\cong\bigcup\limits_{n=0}^{+\infty}\fm_{2^n}(\C)
\label{clifford}
\end{equation}
or equivalently 
\[C(M)\cong\fm_2(\C)\otimes\fm_2(\C)
\otimes\dots\] 
because this injective limit is also isomorphic to the infinite tensor product of 
$\fm_2(\C)$'s. For clarity note that being (\ref{integralas}) a non-local 
operation, $C(M)$ is a genuine global infinite dimensional object.

It is well-known (cf. \cite[Section I.3]{con}) that {\it any}
complexified infinite Clifford algebra like $C(M)$ above generates
{\it the} ${\rm II}_1$-type hyperfinite factor von Neumann algebra. Let us
summarize this procedure too (cf. \cite[Section 1.1.6]{ana-pop}). It readily 
follows that $C(M)$ possesses a unit $1\in C(M)$ and its center comprises the 
scalar multiples of the unit only. Moreover $C(M)$ continues to admit a 
canonical embedding $\Omega^2_c(M;\C)\subset C(M)$ satisfying 
$\omega^2=Q(\omega)1$ where now $Q$ denotes the quadratic form
induced by the complex-bilinear extension of (\ref{integralas}). We also see 
via (\ref{clifford}) already that $C(M)$ is a complex $*$-algebra whose 
$*$-operation (provided by taking Hermitian matrix transpose, a non-local 
operation) is written as $A\mapsto A^\divideontimes$. The isomorphism 
(\ref{clifford}) also shows that if $A\in C(M)$ then one can pick the smallest 
$n\in\N$ such that $A\in\fm_{2^n}(\C)$ consequently $A$ has a finite trace
defined by $\tau (A):=2^{-n}{\rm Trace}(A)$ i.e., taking the usual normalized
trace of the corresponding $2^n\times 2^n$ complex matrix. It is
straightforward that $\tau (A)\in\C$ does not depend on $n$. We can then
define a sesquilinear inner product on $C(M)$ by
$(A,B):=\tau (AB^\divideontimes)$ which is non-degenerate thus the
completion of $C(M)$ with respect to the norm $\Vert\:\cdot\:\Vert$ induced by
$(\:\cdot\:,\:\cdot\:)$ renders $C(M)$ a complex Hilbert space what we shall
write as $\ch$ and its Banach algebra of all bounded linear operators as
$\fb(\ch)$. Multiplication in $C(M)$ from the left on
itself is continuous hence gives rise to a representation $\pi:C(M)\rightarrow
\fb(\ch)$. Finally our central object effortlessly emerges as the weak
closure of the image of $C(M)$ under $\pi$ within $\fb (\ch)$ or
equivalently, by referring to von Neumann's bicommutant theorem
\cite[Theorem 2.1.3]{ana-pop} we put
\[\fr:=(\pi(C(M)))''\subset\fb(\ch)\:\:.\]
This von Neumann algebra of course admits a unit $1\in\fr$ moreover
continues to have trivial center i.e., is a factor. Moreover by
construction it is hyperfinite. The trace $\tau$ as defined extends from
$C(M)$ to $\fr$ and satisfies $\tau (1)=1$. Moreover
\cite[Proposition 4.1.4]{ana-pop} this trace is unique on $\fr$.
Likewise we obtain by extension a representation
$\pi:\fr\rightarrow\fb(\ch)$. The canonical inclusion
$\Omega^2_c(M;\C)\subset C(M)$ recorded above extends to both
$\Omega^2_c(M;\C)\subset\ch$ and $\Omega^2_c(M;\C)\subset\fr$.
Thus in order to carefully distinguish the
two different completions $\fr$ and $\ch$ of one and the same object
$C(M)$ we shall write $A\in\fr$ but $\hat{B}\in\ch$ from now on as usual. 
This is necessary since $\fr$ and $\ch$ are very different for example as
${\rm U}(\ch)$-modules: given a unitary operator $V\in {\rm U}(\ch)$
then $A\in\fr$ is acted upon as $A\mapsto VAV^{-1}$ but $\hat{B}\in\ch$
transforms as $\hat{B}\mapsto V\hat{B}$. Using this notation and introducing
$\hat{A}:=\pi(A)\hat{1}$ the trace always can be written as a scalar product
with the image of the unit in $\ch$ that is, for every $A\in\fr$ we have 
\[\tau(A)=(\hat{A},\hat{1})\]
yielding a general and geometric expression for the trace. 

{\it Exploring the algebra $\fr$.} Before proceeding further let us make 
a digression here to gain a better picture. This is desirable because 
taking the weak limit like $\fr$ of some explicitly known structure like 
$C(M)$ often involves a sort of loosing control over the latter. 
Nevertheless we already know promisingly that $\fr$ is a hyperfinite 
factor von Neumann algebra of ${\rm II}_1$-type. Let us now exhibit some of its 
elements.
 
1. Our first examples are the $2$-forms themselves as it 
follows from the already mentioned canonical embedding $\Omega^2_c(M;\C)
\subset C(M)$ combined with $C(M)\subset\fr$ yielding 
$\Omega^2_c(M;\C)\subset\fr$. Next, proceeding along 
the same lines we can construct an interesting natural continuous embedding
\begin{equation}
i_M:M\longrightarrow\fr
\label{beagyazas}
\end{equation}
of any connected oriented smooth $4$-manifold $M$ into its $\fr$. 
To every sufficiently nice closed subset $\emptyset\subseteqq X\subseteqq M$ 
there is an associated linear subspace $\Omega^2_c(M,X;\C)\subset
\Omega^2_c(M;\C)$ consisting of compactly supported smooth $2$-forms vanishing 
at least along $X$. Furthermore we have seen that as a by-product of the 
Clifford algebra construction there exists a canonical embedding 
$\Omega^2_c(M;\C)\subset\ch$ too. In this way to every point $x\in M$ one
can attach a closed subspace $V_x\subset\ch$ defined by taking the closure
of $\Omega^2_c(M,x\:;\C)$ within $\ch$. Let $Q_x:\ch\rightarrow V_x$ be
the corresponding orthogonal projection. {\it A priori}
$Q_x\in\fb(\ch)$ however in fact $Q_x\in\fr$. This is because $V_x$ arises
as the image of an operator $A_x\in\fr$ which for instance looks like
extending $\omega\mapsto f_x\omega$ with a bounded smooth function $f_x$
vanishing at $x\in M$ from $\Omega^2_c(M;\C)$ to 
$\overline{\Omega^2_c(M;\C)}\subset\ch$ and defined to be 
zero on $\overline{\Omega^2_c(M;\C)}^\perp\subset\ch$. It readily follows
that the resulting map $x\mapsto Q_x$ is injective and
continuous hence gives rise to a continuous embedding. 
To make a comparison, in {\it algebraic geometry} points are
characterized by maximal ideals of an abstractly given commutative ring.
Here the corresponding objects would therefore be the maximal
two-sided (weakly closed) ideals of a von Neumann algebra (regarded as a
non-commutative ring). However in sharp contrast to the commutative situation
a tracial factor von Neumann algebra is always {\it simple} (cf. e.g.
\cite[Proposition 4.1.5]{ana-pop}) consequently in our case the concept of
ideals cannot be used to characterize points hence the reason we used rather
special elements of the von Neumann algebra. (\ref{beagyazas}) is 
a counterpart of embedding Riemannian manifolds into Hilbert spaces via heat 
kernel techniques \cite{ber-bes-gal}.

2. To see more examples, let us return to the Clifford algebra in 
(\ref{clifford}) for a moment. We already know that there exists a 
canonical embedding $\Omega^2_c(M;\C)\subset C(M)$. In addition 
to this let us find a Clifford module for $C(M)$. Consider again any finite 
{\it even} dimensional approximation $C_m(M)=C_{r,s}(M)\otimes\C$ constructed 
from $(V_m,Q_{r,s})$ where now $V_m\subset\Omega^2_c(M)$ is a real even $m=r+s$ 
dimensional subspace. Choose any $2^{\frac{m}{2}}$ dimensional 
complex vector subspace $S_m$ within $\Omega^2_c(M;\C)$. If 
$\ed(\Omega^2_c(M;\C))$ denotes the associative algebra of 
{\it all} $\C$-linear transformations of $\Omega^2_c(M;\C)$ then 
$S_m\subset\Omega^2_c(M;\C)$ induces an embedding 
$\ed S_m\subset\ed(\Omega^2_c(M;\C))$ moreover we know 
that $\ed S_m\cong\fm_{2^\frac{m}{2}}(\C)\cong C_m(M)$. Therefore we obtain a 
non-canonical inclusion $C_m(M)\subset\ed(\Omega^2_c(M;\C))$ for every 
fixed $m\in2\N$. Furthermore 
$S_m\subset S_{m+2}\subset\Omega^2_c(M;\C)$ given by 
$\omega\mapsto\binom{\omega}{0}$ induces a sequence 
$C_m(M)\subset C_{m+2}(M)\subset\ed(\Omega^2_c(M;\C))$ for Clifford algebras 
which is compatible with the previous ascending chain of their matrix 
algebra realizations. Consequently taking the limit 
$m\rightarrow+\infty$ we come up with a non-canonical injective 
linear-algebraic homomorphism 
\begin{equation}
C(M)\subset\ed(\Omega^2_c(M;\C))\:\:.
\label{izomorfizmus}
\end{equation}
Of course, unlike $\Omega^2_c(M;\C)\subset C(M)$ above,  
(\ref{izomorfizmus}) does not exist in the finite dimensional case.

Although the algebra $\ed(\Omega^2_c(M;\C))$ is yet too huge, we can at least 
exhibit some of its elements. The simplest ones are the 
$2$-forms themselves because $\Omega^2_c(M;\C)\subset 
C(M)\subset\ed(\Omega^2_c(M;\C))$ holds as we already 
know.\footnote{The extension of the 
Clifford multiplication $(V_m\otimes\C)\times S_m\rightarrow S_m$ by 
$m\rightarrow +\infty$ thus induces a product structure 
$\Omega^2_c(M;\C)\times\Omega_c^2(M;\C)\rightarrow\Omega^2_c(M;\C)$ rendering 
the space of $2$-forms an infinite dimensional non-unital non-commutative 
associative algebra carrying a non-degenerate symmetric pairing by 
(\ref{integralas}); hence resembling a Frobenius algebra. It is interesting 
that the orientation and smooth structure on $M$ alone induces this structure.} 
Moreover $C^\infty (M;\ed(\wedge^2T^*M\otimes\C))\subset\ed(\Omega^2_c(M;\C))$ 
i.e., bundle morphisms are also included. These are local (algebraic) 
operators but are important because they allow to 
make a contact with local {\it four dimensional} differential 
geometry.\footnote{In fact all the constructions so far work for an arbitrary 
oriented and smooth $4k$-manifold with $k=0,1,2,\dots$ (note that in $4k+2$ 
dimensions the indefinite pairing (\ref{integralas}) gives rise 
to a symplectic structure on $2k+1$-forms).\label{4k}} A 
peculiarity of four dimensions 
is that $C^\infty(M;\ed(\wedge^2T^*M\otimes\C ))$ contains the space of 
curvature tensors on $M$. If $(M,g)$ is an oriented Riemannian $4$-manifold 
then its Riemannian curvature tensor $R_g$ is a member of this 
subalgebra: with respect to the splitting of $2$-forms into their 
(anti)self-dual parts it looks like (cf. \cite{sin-tho})
\begin{equation}
R_g=\left(\begin{matrix}
                \frac{1}{12}{\rm Scal}+{\rm Weyl}^+ & {\rm Ric}_0\\
                  {\rm Ric}_0^* & \frac{1}{12}{\rm Scal}+{\rm Weyl}^-
                      \end{matrix}\right)
\:\:\:\:\::\:\:\:\:\:\begin{matrix}\Omega^+_c(M;\C)\\
                                                  \bigoplus\\
                                                 \Omega^-_c(M;\C)
                                            \end{matrix}
         \:\:\:\longrightarrow\:\:\:\begin{matrix}\Omega^+_c(M;\C)\\
                                                  \bigoplus\\
                                                 \Omega^-_c(M;\C)
                                            \end{matrix}
\label{gorbulet}
\end{equation}
and more generally, $\ed(\Omega^2_c(M;\C))$ contains all algebraic 
(i.e. formal only, not stemming from a metric) curvature tensors $R$ over $M$. 

To see non-local examples, let us say that 
$\omega\in L^1_{loc}(M;\wedge^2T^*M\otimes\C)$ if 
$\langle\omega,\varphi\rangle_{L^2(M)}$ given by (\ref{integralas}) exists 
for all $\varphi\in\Omega^2_c(M;\C)$. Then a tensor field $K$ over $M\times M$ 
is called an {\it admissible double $2$-form over $M$} if the map 
$x\mapsto K(x,y)$ belongs to $C^\infty_c(M;\wedge^2T^*M\otimes\C)=
\Omega^2_c(M;\C)$ for every $y\in M$ while $y\mapsto 
K(x,y)$ gives rise to an element of $L^1_{loc}(M;\wedge^2T^*M\otimes\C)$ for 
every $x\in M$. Picking $R\in C^\infty (M;\ed(\wedge^2T^*M\otimes\C))$ 
the map $\omega\mapsto P\omega$ defined by
\[P\omega(x):=\int\limits_{y\in M}K(x,y)\wedge(R(y)\omega(y))\]
is a non-local pseudo-differential operator acting on $2$-forms and belongs to 
$\ed(\Omega^2_c(M;\C))$. Furthermore orientation-preserving diffeomorphisms 
act on $2$-forms via non-local pullback hence we conclude that 
${\rm Diff}^+(M)\subset\ed(\Omega^2_c(M;\C))$. Note that the 
Lie algebra ${\rm Lie}({\rm Diff}^+(M))\cong C^\infty_c(M;TM)$ consisting of 
compactly supported real vector fields acts $\C$-linearly on $\Omega^2_c(M;\C)$ 
through Lie derivatives hence we also find that 
${\rm Lie}({\rm Diff}^+(M))\subset\ed(\Omega^2_c(M;\C))$. However these 
are local (non-algebraic but differential) operators again. 

How to decide whether or not these elements of $\ed(\Omega^2_c(M;\C))$ 
belong to $\fr$? The key concept here is the trace. Compared with the above 
trace expression generally valid on $\fr$, more 
specific trace formulata are obtained if $M$ is endowed with a normalized 
Riemannian metric $g$ i.e., the corresponding volume form $\mu_g=*1$ 
satisfies $\int_M\mu_g=1$. The unique sesquilinear extension 
of $g$ induces a positive definite sesquilinear $L^2$-scalar product 
\[(\varphi, \psi)_{L^2(M,g)}:=\int\limits_{x\in M}g(\varphi(x),\psi(x))\mu_g(x)
=\int\limits_M\varphi\wedge *\overline{\psi}\] 
on $\Omega^2_c(M;\C)$. If $\{\varphi_1,\varphi_2,\dots\}$ is a smooth 
orthonormal frame in $\Omega^2_c(M;\C)$ then it readily follows that 
the trace of any $B\in\ed(\Omega^2_c(M;\C))$ formally looks like 
\[\tau (B)=\lim\limits_{n\rightarrow+\infty}\frac{1}{2^n}\sum\limits_{i=1}^{2^n}
(B\varphi_i\:,\:\varphi_i)_{L^2(M,g)}\]
and, if exists, is independent of the frame used. Obviously 
$B\in\ed(\Omega^2_c(M;\C))\cap C(M)$ if and only if the 
sum on the right hand side is constant after finitely many terms; and an 
inspection of this trace expression shows that in general 
$B\in\ed(\Omega^2_c(M;\C))\cap\fr$ if and only if 
$\tau (B)$ exists. As a consequence note that $\ed(\Omega^2_c(M;\C))\cap\fr$ 
is already independent of the particular inclusion (\ref{izomorfizmus}). 
An example: any $\Phi\in{\rm Diff}^+(M)$ acts on 
$\ed(\Omega^2_c(M;\C))$ by conjugation hence using (\ref{izomorfizmus}) 
extends to a unitary operator on $\ch$ yielding $\vert\tau (\Phi)\vert=1$ 
consequently we obtain ${\rm Diff}^+(M)\subset\ed(\Omega^2_c(M;\C))\cap\fr$.

When $R\in C^\infty (M;\ed (\wedge^2T^*M\otimes\C))$ the previous trace 
formula can be further specified 
because one can compare the global trace $\tau(R)$ and the local trace function 
$x\mapsto {\rm tr}(R(x))$ given by the pointwise traces of the local operators 
$R(x):\wedge^2T^*_xM\otimes\C\rightarrow\wedge^2T^*_xM\otimes\C$ at every 
$x\in M$. Recall that $\fr$ has been constructed 
as the weak closure of the Clifford algebra (\ref{clifford}). In fact 
\cite[Section 1.1.6]{ana-pop} the 
universality of $\fr$ permits to obtain it from other matrix algebras too, 
like for instance from $\bigcup\limits_{n=0}^{+\infty}\fm_{6^n}(\C)$ whose weak 
closure therefore is again $\fr$. By the aid of this altered construction 
we can formally start with 
\[\tau (R)=\lim\limits_{n\rightarrow+\infty}\frac{1}{6^n}\sum\limits_{i=1}^{6^n}
(R\varphi_i\:,\:\varphi_i)_{L^2(M,g)}\:\:.\]
Fix $n\in\N$, write $M_n:=\bigcap\limits_{i=1}^{6^n}
{\rm supp}\varphi_i\subseteqq M$ and take a point $x\in M_n$. 
Since $\dim_\C(\wedge^2T^*_xM\otimes\C)=\binom{4}{2}=6$ the maximal number of 
completely disjoint linearly independent sub-$6$-tuples in 
$\{\varphi_1(x),\varphi_2(x),\dots,\varphi_{6^n}(x)\}$ is equal to 
$\frac{6^n}{6}=6^{n-1}$. Moreover it follows from Sard's lemma that with a 
generic smooth choice for $\{\varphi_1,\varphi_2,\dots\}$ the 
subset of those points $y\in M_n$ where this number is less than $6^{n-1}$ 
has measure zero in $M_n$ with respect to the measure $\mu_g$. Consequently
\[\sum\limits_{i=1}^{6^n}(R\varphi_i,\varphi_i)_{L^2(M_n,g)}=
\int\limits_{x\in M_n}\sum\limits_{i=1}^{6^n}g(R(x)\varphi_i(x),\varphi_i(x))
\:\mu_g(x)=6^{n-1}\int\limits_{x\in M_n}{\rm tr} (R(x))\mu_g(x)\:\:.\]
Since $\{\varphi_1,\varphi_2,\dots\}$ is a basis in 
$\Omega^2_c(M;\C)$ therefore $M\setminus\bigcup\limits_{n=0}^{+\infty}M_n$ 
has measure zero as well we can let $n\rightarrow+\infty$ to end up with 
\begin{equation}
\tau (R)=\frac{1}{6}\int\limits_M{\rm tr}(R)\mu_g
\label{nyomok}
\end{equation}
and again $R\in C^\infty (M;\ed (\wedge^2T^*M\otimes\C))\cap\fr$ if 
and only if (\ref{nyomok}) exists. Also note that $\tau$ is in fact the 
generalization of the total scalar curvature of a 
Riemannian manifold. So taking into account (\ref{izomorfizmus}) 
we obtain useful criteria for checking whether or not an operator in 
$\ed(\Omega^2_c(M;\C))$ belongs to the completion of $C(M)$ which is 
$\fr$. For instance if the curvature 
$R_g$ of $(M,g)$ as an operator in (\ref{gorbulet}) is bounded which means that
\[\sup\limits_{\Vert\omega\Vert_{L^2(M,g)}=1}
\Vert R_g\omega\Vert_{L^2(M,g)}\leqq K<+\infty\]
then 
\[0\leqq\vert\tau (R_g)\vert\leqq
\lim\limits_{n\rightarrow+\infty}\frac{1}{2^n}\sum\limits_{i=1}^{2^n}
\vert(R_g\varphi_i\:,\:\varphi_i)_{L^2(M,g)}\vert\leqq
\lim\limits_{n\rightarrow+\infty}\frac{1}{2^n}\sum\limits_{i=1}^{2^n}
\Vert R_g\varphi_i\Vert_{L^2(M,g)}\leqq K\]
yielding $R_g\in C^\infty (M;\ed (\wedge^2T^*M\otimes\C))\cap\fr$ and 
more generally $R\in C^\infty (M;\ed (\wedge^2T^*M\otimes\C))
\cap\fr$ if $R$ is bounded. Similar condition stems from (\ref{nyomok}). 

3. We close the partial comprehension of $\fr$ with an observation regarding 
its overall structure. The canonical inclusion $\Omega^2_c(M;\C)\subset\fr$ 
(and likewise $\Omega^2_c(M;\C)\subset\ch$) implies that $\fr$ (and likewise 
$\ch$) is generated by all finite products $\omega_1\omega_2\dots\omega_k$ 
of $2$-forms within $C(M)$ however this is not very 
informative. Rather consider the well-known short exact sequence of groups 
involving the fiberwise automorphism group (the gauge group) and the global 
automorphism group of the vector bundle $\wedge^2T^*M\otimes\C$ as well as the 
orientation-presering diffeomorphism group of the underlying $M$ respectively: 
\[1\longrightarrow C^\infty (M;{\rm Aut}(\wedge^2T^*M\otimes\C))
\longrightarrow {\rm Aut}(\Omega^2_c(M;\C))
\longrightarrow {\rm Diff}^+(M)\longrightarrow 1\]
whose shape at the Lie algebra level looks like
\[0 \longrightarrow C^\infty (M;{\rm End}(\wedge^2T^*M\otimes\C))
\longrightarrow{\rm End}(\Omega^2_c(M;\C))
\longrightarrow {\rm Lie}({\rm Diff}^+(M))\longrightarrow 0\:\:.\]
We already have an embedding (\ref{izomorfizmus}). In addition to this 
there exists an isomorphism of Lie algebras 
$L:C^\infty_c(M;TM)\rightarrow {\rm Lie}({\rm Diff}^+(M))$ such that 
$X\mapsto L_X$ is nothing but 
taking Lie derivative with respect to a compactly supported real vector 
field where the first-order $\C$-linear differential operator $L_X$ is 
supposed to act on $2$-forms hence ${\rm Lie}({\rm Diff}^+(M))
\subset \ed(\Omega^2_c(M;\C))$ as we know already too. 
Therefore the intersection of the latter sequence with $C(M)$ is meaningful and 
gives
\[0 \longrightarrow C^\infty (M;{\rm End}(\wedge^2T^*M\otimes\C))\cap C(M)
\longrightarrow C(M)\longrightarrow 
{\rm Lie}({\rm Diff}^+(M))\cap C(M) \longrightarrow 0\:\:.\]
The second term consists of fiberwise algebraic hence local operators 
having finite trace (\ref{nyomok}) and likewise the fourth terms 
consists of finite trace Lie derivatives thus 
belongs to the class of local operators too. Since the vector 
spaces underlying $C(M)$ considered either as an associative or a 
Lie algebra are isomorphic we conclude, as an important 
structural observation, that the overall construction here is geometric in 
the sense that {\it the algebra $\fr$ is generated by (trace class) local 
operators}. 

Summing up all of our findings so far: $\fr$ is a hyperfinite factor von 
Neumann algebra of ${\rm II}_1$-type associated to $M$ such that the solely 
input in its construction has been the pairing (\ref{integralas}). 
Hence $\fr$ depends only on the orientation and the smooth structure of $M$ 
in a functorial way. It contains, certainly among many other non-local 
operators, the space $M$ itself via (\ref{beagyazas}), its 
orientation-preseving diffeomorphisms as well as its space of bounded 
algebraic curvature tensors. Nevertheless $\fr$ is geometric in the sense that 
it is generated by $M$'s finite trace local operators alone. It is 
remarkable that despite the plethora of smooth $4$-manifolds detected since 
the 1980's their associated von Neumann algebras here are unique up to 
isomorphisms 
of von Neumann algebras (cf. e.g. \cite[Theorem 11.2.2]{ana-pop}) offering 
a sort of justification terming $\fr$ as ``universal''. One is therefore 
tempted to look upon $\fr$ as a natural common non-commutative space 
generalization of either all the pure oriented smooth $4$-manifolds $M$ 
or rather all the (pseudo-)Riemannian ones $(M,g)$. This universality also 
justifies the simple notation $\fr$ used throughout the text.

{\it A new smooth $4$-manifold invariant.} After these preliminary 
considerations we are in a position to approach four dimensional smoothness 
from a new operator-algebraic direction.
 
\begin{lemma} Let $M$ be a connected oriented smooth $4$-manifold 
and $\fr$ its von Neumann algebra with trace $\tau$ as before. Then there 
exists a complex separable Hilbert space $\ci (M)^\perp$ and a representation 
$\rho_M:\fr\rightarrow\fb (\ci (M)^\perp)$ with the following 
properties. If $\pi:\fr\rightarrow\fb (\ch)$ is the 
representation constructed above then 
$\{0\}\subseteqq\ci (M)^\perp\subsetneqq\ch$ and 
$\rho_M=\pi\vert_{\ci (M)^\perp}$ holds; therefore, although 
$\rho_M$ can be the trivial representation, it is surely not unitary 
equivalent to the standard representation. Moreover the unitary 
equivalence class of $\rho_M$ is invariant under orientation-preserving 
diffeomorphisms of $M$. 

Thus the Murray--von Neumann coupling constant\footnote{Also 
called the {\it $\fr$-dimension} of a left $\fr$-module hence 
denoted $\dim_{\fr}$, cf. \cite[Chapter 8]{ana-pop}.\label{dimenzio}} 
of $\rho_M$ is invariant under orientation-preserving diffeomorphisms. 
Writing $P_M:\ch\rightarrow\ci(M)^\perp$ for the orthogonal 
projection the coupling constant is equal to $\tau (P_M)\in [0,1)\subset\R_+$ 
consequently $\gamma (M):=\tau (P_M)$ is a smooth $4$-manifold invariant.
\label{terbelilemma}
\end{lemma} 

\begin{proof} First let us exhibit a representation of $\fr$; 
this construction is inspired by the general Gelfand--Naimark--Segal 
technique however exploits the special features of our construction so far as 
well. Pick a pair $(\Sigma ,\omega)$ consisting of an (immersed) closed 
orientable surface $\Sigma\looparrowright M$ with induced oriantation 
and $\omega\in\Omega^2_c(M;\C)$ which is also closed i.e., $\dd\omega=0$. 
Consider the differential geometric $\C$-linear functional 
$F_{\Sigma,\omega}:\fr\rightarrow\C$ by continuously extending 
\[A\longmapsto\int\limits_\Sigma A\omega\] 
from $\ed(\Omega^2_c(M;\C))\cap\fr$. This extension is unique because 
$\ed(\Omega^2_c(M;\C))\cap\fr$ is norm-dense in $\fr$. In case of 
$F_{\Sigma ,\omega}(1)\not=0$ let $\{0\}\subseteqq I(M)
\subseteqq\fr$ be the subset of elements $A\in\fr$ 
satisfying $F_{\Sigma,\omega}(A^\divideontimes A)=0$. In fact for 
all pairs $(\Sigma,\omega)$ we obviously find 
$\{0\}\subsetneqq I(M)$ and $I(M)\cap\C1=\{0\}$ hence $I(M)\subsetneqq\fr$ too. 
We assert that $I(M)$ is a multiplicative left-ideal 
in $\fr$ which is independent of $(\Sigma,\omega)$. In the case of 
$F_{\Sigma ,\omega}(1)=0$ we put $I(M)=\fr$ hence it is again independent 
of $(\Sigma,\omega)$ but trivially in this way. 

Consider the case when $F_{\Sigma, \omega}(1)\not=0$. Then we can 
assume that $F_{\Sigma,\omega}(1)=1$ hence $F_{\Sigma, \omega}$ is a 
positive functional; applications of the standard inequality 
$\vert F_{\Sigma,\omega}(A^\divideontimes B)\vert^2\leqq 
F_{\Sigma,\omega}(A^\divideontimes A)F_{\Sigma,\omega}(B^\divideontimes B)$ 
show that $I_{\Sigma ,\omega}(M)$ defined by the elements satisfying 
$F_{\Sigma,\omega}(A^\divideontimes A)=0$ is a multiplicative left-ideal in 
$\fr$. Concerning its $\omega$-dependence, let $\omega'\in\Omega^2_c(M;\C)$ 
be another closed $2$-form having the property $F_{\Sigma,\omega'}(1)=1$; 
since neither $\omega$ nor $\omega'$ are identically zero, 
we can pick an {\it invertible} 
operator $T\in{\rm Aut}(\Omega^2_c(M;\C))$ satisfying 
$\omega '\vert_\Sigma =T\omega\vert_\Sigma$. Then by 
$F_{\Sigma,\omega '}(A^\divideontimes A)=
F_{\Sigma,\omega}(A^\divideontimes AT)$ and applying the above inequality 
in the form 
$\vert F_{\Sigma ,\omega'}(A^\divideontimes AT)\vert^2\leqq F_{\Sigma,\omega} 
(A^\divideontimes A)F_{\Sigma ,\omega}((AT)^\divideontimes (AT))$ we find 
$I_{\Sigma,\omega'}(M)\subseteqq I_{\Sigma,\omega}(M)$. Likewise, making use of 
$F_{\Sigma,\omega}(A^\divideontimes A)=
F_{\Sigma ,\omega'}(A^\divideontimes AT^{-1})$ and 
$\vert F_{\Sigma ,\omega}(A^\divideontimes AT^{-1})\vert^2\leqq F_{\Sigma 
,\omega'}(A^\divideontimes A)F_{\Sigma ,\omega'}((AT^{-1})^\divideontimes 
(AT^{-1}))$ imply the converse inequality 
$I_{\Sigma ,\omega'}(M)\supseteqq I_{\Sigma ,\omega}(M)$. 
Consequently $I_{\Sigma ,\omega'}(M)=I_{\Sigma ,\omega}(M)$.  

Concerning the general $(\Sigma,\omega)$-dependence of $I_{\Sigma,\omega}$ 
we argue as follows. Let $\eta_\Sigma\in\Omega^2(M;\R)$ be a closed real 
$2$-form representing the Poincar\'e-dual $[\eta_\Sigma ]\in H^2(M;\R )$ of 
$\Sigma\looparrowright M$. Equip $M$ with an arbitrary Riemannian metric $g$; 
since the corresponding Hodge operator $*$ is an isomorphism on $\wedge^2T^*M$ 
we can take a $2$-form $\varphi_\Sigma$ such that 
$\eta_\Sigma =*\varphi_\Sigma$. Then via 
$\int_\Sigma\omega =\int_M\omega\wedge\overline{\eta}_\Sigma =
\int_M\omega\wedge*\overline{\varphi}_\Sigma=
(\omega,\varphi_\Sigma)_{L^2(M,g)}$ 
the functional can be re-expressed as $F_{\Sigma,\omega}(A^\divideontimes A)= 
(A^\divideontimes A\omega,\varphi_\Sigma)_{L^2(M,g)}$ in terms of the 
corresponding definite sesquilinear $L^2$-scalar product on $(M,g)$. 
Let $\Sigma '\looparrowright M$ be another closed 
surface and $\omega'$ another closed $2$-form such that 
$F_{\Sigma ',\omega'}(1)=1$. Altering $\omega$ and $\omega'$ along 
$\Sigma$ and $\Sigma'$ respectively if necessary 
(which has no effect on $I(M)$ as we have seen) we can pick some 
compactly supported $2$-form $\Omega$ on $M$ such 
that $\Omega\vert_{\Sigma}=\omega$ and $\Omega\vert_{\Sigma'}=\omega'$. 
Moreover as before take a representative 
$*\varphi_{\Sigma '}\in\Omega^2(M;\R )$ for the Poincar\'e-dual of $\Sigma'$. 
We can use again $T\in{\rm Aut}(\Omega^2_c(M;\C))$ 
satisfying $\varphi_{\Sigma '}=T\varphi_\Sigma$. Thus
\[F_{\Sigma',\omega'}(A^\divideontimes A)
=(A^\divideontimes A\Omega ,\varphi_{\Sigma'})_{L^2(M,g)}=
(A^\divideontimes A\Omega,T\varphi_\Sigma)_{L^2(M,g)}=(T^\divideontimes
A^\divideontimes A\Omega,\varphi_\Sigma)_{L^2(M,g)}=
F_{\Sigma ,\omega}(T^\divideontimes A^\divideontimes A)\] 
together with $\vert F_{\Sigma,\omega}(T^\divideontimes A^\divideontimes A)
\vert^2\leqq F_{\Sigma ,\omega}\big((AT)^\divideontimes AT\big)
F_{\Sigma ,\omega}(A^\divideontimes A)$ 
demonstrate that $I_{\Sigma ',\omega'}(M)\subseteqq I_{\Sigma ,\omega}(M)$. 
In the same fashion $F_{\Sigma ,\omega}(A^\divideontimes A)=F_{\Sigma',\omega'}
((T^{-1})^\divideontimes A^\divideontimes A)$ together with the corresponding 
inequality convinces us that 
$I_{\Sigma',\omega'}(M)\supseteqq I_{\Sigma,\omega}(M)$. 
Thus $I_{\Sigma',\omega'}(M)=I_{\Sigma,\omega}(M)$. 

Secondly if $(\Sigma,\omega)$ is such that $F_{\Sigma,\omega}(1)=0$ 
then by definition $I_{\Sigma,\omega}(M)=\fr$. Therefore if 
$(\Sigma',\omega')$ is another pair with $F_{\Sigma',\omega'}(1)=0$ then 
$I_{\Sigma,\omega}(M)=I_{\Sigma',\omega'}(M)$ (and equal to $\fr$). We 
are now convinced that it is correct to write 
$I_{\Sigma,\omega}(M)$ as $I(M)$. In summary it 
satisfies $\{0\}\subsetneqq I(M)\subseteqq \fr$.

Let us proceed further by exploiting now the observation made during the 
construction of $\fr$ that it acts on a Hilbert space $\ch$ 
with scalar product $(\:\cdot\:,\:\cdot\:)$ by the 
representation $\pi$ i.e., multiplication from the left. In fact, since 
$(\hat{A},\hat{B})=\tau(AB^\divideontimes)$ we see that $\pi$ is the standard 
representation. Consider the space $\{0\}\subsetneqq 
I(M)I(M)^\divideontimes\subseteqq\fr$ 
consisting of all finite sums $A_1B_1+\dots+A_kB_k\in\fr$ where 
$A_i\in I(M)$ and similarly $B_j\in I(M)^\divideontimes$. It gives rise to a 
closed linear subspace $\{0\}\subsetneqq\ci(M)\subseteqq\ch$ 
by taking the closure of $C(M)\cap I(M)I(M)^\divideontimes$ within 
$\ch\supset C(M)$. Therefore $\ci(M)$ is a well-defined closed 
subspace of $\ch$ which is non-trivial if $F_{\Sigma,\omega}(1)\not=0$ 
and coincides with $\ch$ whenever $F_{\Sigma,\omega}(1)=0$. Take its 
orthogonal complementum $\{0\}\subseteqq\ci(M)^\perp\subsetneqq\ch$. Note that 
$\ci (M)^\perp$ is isomorphic to $\ch/\ci(M)$. Taking into account that the 
subset $I(M)$ is a multiplicative left-ideal $\pi:\fr\rightarrow\fb (\ch)$ 
given by left-multiplication restricts to $\ci(M)$ but even more, since the 
scalar product on $\ch$ satisfies the identity 
$(\widehat{AB},\widehat{C}\:)=(\hat{B},\widehat{A^\divideontimes C})$ the 
standard representation restricts to $\ci(M)^\perp$ as well. This latter 
representation is either a unique non-trivial representation if 
$\ci (M)^\perp\not=\{0\}$ (provided by a functional over $M$ with 
$F_{\Sigma,\omega} (1)\not=0$ if exists), or the 
trivial one if $\ci (M)^\perp=\{0\}$ (provided by a functional with 
$F_{\Sigma,\omega}(1)=0$ which always exists). Keeping these in 
mind, for a given $M$ we define 
\[\rho_M:\fr\rightarrow\fb (\ci(M)^\perp)\:\:\:\mbox{to be}\:\:\:
\left\{\begin{array}{ll}
\mbox{$\pi\vert_{\ci (M)^\perp}$ on $\ci (M)^\perp\not=\{0\}$ if possible,}\\ 
\mbox{$\pi\vert_{\ci (M)^\perp}$ on $\ci (M)^\perp=\{0\}$ otherwise.}
\end{array}\right.\]
The choice is unambigously determined by the topology of $M$ 
(see the {\it Remark} below).

From the general theory \cite[Chapter 8]{ana-pop} we know that if 
$P_M:\ch\rightarrow\ci (M)^\perp$ is the orthogonal projection 
then $P_M\in\fr$ because $\ch$ is the standard 
$\fr$-module. The Murray--von Neumann coupling constant of $\rho_M$ depends only
on the unitary equivalence class of $\rho_M$ and is equal to 
$\tau (P_M)\in[0,1]$. However observing that $\rho_M$ is surely not 
isomorphic to $\pi$ since $\ci(M)$ is never trivial the case $\tau (P_M)=1$ 
is excluded i.e., $\tau (P_M)\in [0,1)$. Let $\Phi :M\rightarrow M$ 
be an orientation-preserving diffeomorphism. It induces an inner automorphism 
$A\mapsto\Phi^* A(\Phi^*)^{-1}$ of $\fr$. Taking into account 
that the scalar product on $\ch$ is induced by the trace which is invariant 
under cyclic permutations this inner automorphism is unitary. 
Moreover it transforms $I_{\Sigma ,\omega}(M)$ into 
$I_{\Sigma',\omega'}(M)=I_{\Phi(\Sigma ),\Phi^*\omega}(M)$ hence 
$F_{\Sigma,\omega}(1)=0$ if and only if $F_{\Sigma',\omega'}(1)=0$ 
consequently the Hilbert space $\ci (M)^\perp$ is invariant under $\Phi$. 
Thus $\Phi$ transforms $\rho_M$ into a new representation 
$\Phi^*\rho_M (\Phi^*)^{-1}$ on $\ci (M)^\perp$ which is unitary 
equivalent to $\rho_M$. 

We conclude that $\gamma (M):=\tau (P_M)\in [0,1)$ 
is a smooth invariant of $M$ as stated. 
\end{proof}

\begin{remark}\rm Note that $\gamma (M)=0$ corresponds to the situation when 
$\rho_M$ is the trivial representation on $\ci (M)^\perp=\{0\}$. 
To avoid this we have to demand $F_{\Sigma,\omega}(1)\not=0$ which by the 
closedness assumptions on $\Sigma\looparrowright M$ and 
$\omega\in\Omega^2_c(M;\C)$ is in fact a {\it topological 
condition}: it is equivalent that 
\[\frac{1}{2\pi\sqrt{-1}}F_{\Sigma,\omega}(1)=
\frac{1}{2\pi\sqrt{-1}}\int\limits_\Sigma\omega 
=\langle [\Sigma],[\omega]\rangle\in\C\] 
as a pairing of $[\Sigma]\in H_2(M;\Z)$ and $[\omega]\in H^2(M;\C)$ in 
homology is not trivial. Hence $\gamma (M)=0$ iff 
$H_2(M;\C)=H_2(M;\Z)\otimes\C=\{0\}$ (or equivalently, 
$H^2(M;\C)=H^2(M;\Z)\otimes\C=\{0\}$). Thus unfortunately $\gamma (M)=0$ for 
all acyclic or aspherical manifolds (including homology $4$-spheres). 
Examples in the simply connected case are $M=S^4,\R^4, R^4$ (this latter is 
any exotic or fake $\R^4$) while $M=S^3\times S^1$ is a non-simply 
connected example.  
\end{remark}

\noindent Jones' subfactor theory (for a summary cf. e.g. 
\cite[Section 9.4]{ana-pop} or \cite[Chapter V.10]{con}) 
imposes an interesting restriction on the possible spectrum of $\gamma$ just 
constructed. 

\begin{lemma} Let $M$ be a connected oriented smooth $4$-manifold and 
$\gamma (M)\in[0,1)$ its smooth invariant. Then $\gamma (M)=1-\frac{1}{x}$ 
where $x\in\{4\cos^2\big(\frac{\pi}{n}\big)\:\vert\:n\geqq 3\}\cup[4,+\infty)$ 
that is, an element from the set of all possible Jones' subfactor indices. 
\label{jones}
\end{lemma}
 
\begin{proof} Taking into account that $\{0\}\subsetneqq 
I(M)I(M)^\divideontimes\subseteqq\fr$ is self-adjoint it generates a 
von Neumann subalgebra; more precisely put 
\[\fii (M):=(\pi(C(M)\cap I(M)I(M)^\divideontimes))''
\subset\fb(\ch)\:\:.\] 
Since the set $C(M)\cap I(M)I(M)^\divideontimes$ is already weakly dense in 
$I(M)I(M)^\divideontimes$ which is a non-trivial left-ideal of 
$\fr$ its only defect to be a von Neumann algebra is that 
$C(M)\cap I(M)I(M)^\divideontimes$ is not weakly closed and 
possesses no unit; therefore its weak closure $\fii (M)$ is 
a factor. Thus it is a subfactor of $\fr$ therefore it possesses a 
corresponding Jones index denoted as usual by $[\fr:\fii (M)]\in
\{4\cos^2\big(\frac{\pi}{n}\big)\:\vert\:n\geqq 3\}\cup[4,+\infty)$\:.

The left-ideal $\{0\}\subsetneqq C(M)\cap I(M)I(M)^\divideontimes
\subseteqq C(M)$ acts on itself by letf-multiplications. Since both the von 
Neumann algebra $\C\cong Z(\fr)\subsetneqq\fii (M)\subseteqq\fr$ and the 
Hilbert space $\{0\}\subsetneqq\ci (M)\subseteqq\ch$ 
arise as appropriate closures of this same left-ideal its left 
action on itself extends to a representation of $\fii (M)$ on $\ci(M)$ which, 
taking into account that $\ci (M)\not=\{0\}$, is equivalent to the 
standard representation of $\fii (M)$. We also know already that 
$\ch=\ci(M)\oplus\ci(M)^\perp$ as 
$\fr$-modules by Lemma \ref{terbelilemma}. Recalling the basic properties 
of the dimension function of a left von Neumann algebra module over the von 
Neumann algebra itself (cf. e.g. \cite[Chapter 8]{ana-pop}) we collect:  
\begin{eqnarray}
\dim_{\fr}\ch&=&\dim_{\fr}\ci(M)+
\dim_{\fr}\ci(M)^\perp\:\:\:\:\:\mbox{(additivity)}\nonumber\\
\dim_{\fr}\ch&=&1\:\:\:\:\:\mbox{(the standard representation of
$\fr$)}\nonumber\\
\dim_{\fr}\ci(M)^\perp&=&\gamma (M)\:\:\:\:\:\mbox{(by 
Lemma \ref{terbelilemma} and Footnote \ref{dimenzio})} \nonumber\\
\dim_{\fii (M)}\ci (M)&=&[\fr\::\:\fii (M)]\dim_{\fr}\ci(M)
\:\:\:\:\:\mbox{(dimension comparison)}\nonumber\\
\dim_{\fii (M)}\ci (M)&=&1 \:\:\:\:\:
\mbox{(the standard representation of $\fii(M)$)}\nonumber
\end{eqnarray}
from which it follows that $\gamma (M)=1-\frac{1}{[\fr\::\:\fii (M)]}$ 
yielding the desired result. 
\end{proof}

\noindent {\it Proof of Theorem \ref{nagytetel1}}. This theorem follows from 
the general construction of $\fr$ as we outlined here and in particular 
from Lemmata \ref{terbelilemma} and \ref{jones}.\hspace{10.3cm}$\square$ 
\vspace{0.1in}

\noindent Next we collect some basic useful properties of the invariant.

\begin{lemma} (Reversing orientation.) If $M$ is a connected oriented smooth 
$4$-manifold and $\overline{M}$ is its orientation-reversed form then 
$\gamma (\overline{M})=\gamma (M)$.

(Excision principle.) Let $M$ be a connected oriented 
smooth $4$-manifold and $\emptyset\subseteqq 
Y\subset M$ a submanifold so that $M\setminus Y\subseteqq M$ 
is connected and the embedding $i: M\setminus Y\rightarrow M$ induces an 
isomorphism $i_*: H_2(M\setminus Y;\Z)\rightarrow H_2(M;\Z)$ on the 
$2^{\rm nd}$ homology. Then $M\setminus Y$ with induced orientation and 
smooth structure is a connected oriented smooth $4$-manifold satisfying 
$\gamma (M\setminus Y)=\gamma (M)$.

(Gluing principle.) Let $M$ and $N$ be two connected, oriented smooth 
$4$-manifolds and write $M\#N$ for their connected sum. With induced 
orientation $M\#N$ is a connected, oriented smooth $4$-manifold. Its 
smooth invariant satisfies 
\[\gamma (M\#N)=\frac{\gamma (M)+\gamma (N)}{1+\gamma (M)\gamma (N)}\:\:\:.\] 
\label{gammaformulak}
\end{lemma}

\begin{proof} The first assertion is obvious from 
$\gamma$'s construction carried out in the proof of Lemma \ref{terbelilemma}. 

Regarding the second assertion $M\setminus Y$ is a 
connected oriented smooth $4$-manifold by assumption consequently admits an 
associated von Neumann algebra $\fr (M\setminus Y)$ which is also a 
hyperfinite ${\rm II}_1$ factor. The space of compactly supported smooth 
forms on $M\setminus Y$ can be identified with the linear subspace 
consisting of compactly supported smooth forms on $M$ which vanish along 
$Y\subset M$ that is, $\Omega^2_c(M\setminus Y;\C)\cong
\Omega^2_c(M,Y;\C)\subset\Omega^2_c(M;\C)$. By the aid of this embedding it 
is easy to see that in the $L^2$-topology induced by any Riemannian metric 
$\Omega^2_c(M\setminus Y;\C)$ is dense in $\Omega^2_c(M;\C)$ 
which immediately implies that $C(M\setminus Y)=C(M)$ for the corresponding 
complexified Clifford algebras; thus taking closures we come up with 
$\ch(M\setminus Y)=\ch(M)$ and then $\fr(M\setminus Y)=\fr(M)$. By definition 
$\fr (M\setminus Y)$ acts via $\rho_{M\setminus Y}$ on 
$\ci (M\setminus Y)^\perp$ and $\fr (M)$ acts via 
$\rho_M$ on $\ci (M)^\perp$. By assumption $M\setminus Y\subseteqq M$ induces 
an isomorphism on the $2^{\rm nd}$ homology hence $\ci (M\setminus Y)^\perp$ 
and $\ci (M)^\perp$ are simultaneously trivial or not. In fact 
$\ci (M\setminus Y)^\perp=\ci (M)^\perp$. This is because recall that 
the condition $A\in I(M\setminus Y)$ can be written as 
$F_{\Sigma, \varphi_\Sigma}(A^\divideontimes A)=
(A^\divideontimes A\varphi_\Sigma, \varphi_\Sigma)_{L^2(M\setminus Y,g)}=0$ 
where $\varphi_\Sigma$ is the Hodge-dual of the harmonic representative 
$\eta_\Sigma$of the Poincar\'e-dual of $\Sigma\looparrowright M\setminus Y$; 
since $Y$ has zero $4$ dimensional Lebesgue measure $A\in I(M\setminus Y)$ if 
and only if $A\in I(M)$; therefore $\ci (M\setminus Y)=\ci (M)$ yielding 
the same for the complementums in $\ch (M\setminus Y)=\ch (M)$. 
Consequently both $\rho_{M\setminus Y}$ and $\rho_M$ are representations of 
$\fr(M\setminus Y)=\fr (M)$ on $\ci(M\setminus Y)^\perp=\ci(M)^\perp$. 
By definition $\gamma (M\setminus Y)=
\dim_{\fr (M\setminus Y)}\ci(M\setminus Y)^\perp$ 
and likewise $\gamma (M)=\dim_{\fr (M)}\ci(M)^\perp$. 
Therefore $\gamma (M\setminus Y)=\gamma (M)$ as stated. 

Concerning the third assertion note that the $\gamma$-invariant 
is a well-defined map from (the category) $\sm$ of all 
orientation-preserving diffeomorphism classes of connected, oriented smooth 
$4$-manifolds into the real interval $[0,1)\subset\R$. But $\sm$ 
forms a commutative semigroup with unit $S^4$ under the connected sum 
operation $\#$. That is, if $X,Y,Z\in\sm$ and $S^4\in\sm$ is the 
$4$-sphere then $X\#Y\cong Y\#X$ and $(X\#Y)\#Z\cong X\#(Y\#Z)$ and 
$X\#S^4\cong X$. Pick $M,N\in\sm$ with their connected sum $M\#N\in\sm$ 
and consider the corresponding invariants $\gamma (M),\gamma (N),\gamma 
(M\#N)\in [0,1)$. Define $\bullet :[0,1)\times [0,1)\rightarrow [0,1)$ 
by setting $\gamma (M)\bullet\gamma (N):=\gamma(M\#N)$. 
The $\bullet$-operation is therefore well-defined and has the properties 
$\gamma (X)\bullet\gamma 
(Y)=\gamma (Y)\bullet\gamma (X)$ and $(\gamma (X)\bullet\gamma 
(Y))\bullet\gamma (Z)=\gamma (X)\bullet (\gamma
(Y)\bullet\gamma (Z))$ and $\gamma (X)\bullet\gamma (S^4)=\gamma (X)$.  
These ensure us that $([0,1),\bullet )$ is a unital commutative semigroup 
and $\gamma :(\sm,\#)\rightarrow ([0,1),\bullet )$ is a unital semigroup 
homomorphism. Mapping the unique semigroup structure of 
$[0,+\infty)\subset\R$ to $[0,1)$ by 
$\alpha\mapsto{\rm tanh}\:\alpha$ gives $x\bullet y=\frac{x+y}{1+xy}$ 
with $0$ being the unit. Thus $\gamma (S^4)=0$ (as we already know) but 
otherwise $\gamma$ is not trivial because $\gamma (M)>0$ if $b_2(M)>0$. This 
yields the shape for $\gamma (M)\bullet\gamma (N)$ as stated. 
\end{proof}

\noindent {\it Proof of Theorem \ref{nagytetel2}.} This theorem follows from 
Lemma \ref{gammaformulak}.\hspace{6.5cm}$\square$
\vspace{0.01in}

\noindent More generally one can demonstrate by techniques 
from $4$-manifold theory and the gluing principle the following non-trivial but 
quite insensitive behaviour of $\gamma$ in the closed simply connected case. 

\begin{lemma}
If $M'$ and $M''$ are connected, simply connected, closed 
smooth $4$-manifolds which are homeomorphic then $\gamma (M')=\gamma (M'')$. 

Take $y\in [0,1)$ and the sequence
\[R_0(y):=0\:,\:R_1(y):=y\:,\:\dots\:,\: R_k(y):=\frac{y+R_{k-1}(y)}
{1+yR_{k-1}(y)}\:,\:\dots\]
for all $k=0,1,2,\dots$ representing the semigroup
$\{0\}\cup\N$ inside $[0,1)$. Then for any connected, simply connected,
closed smooth $4$-manifold 
\[\gamma (M)=R_{b_2(M)}\left(\begin{smallmatrix}\frac{8}{9}
\end{smallmatrix}\right)=
\frac{17^{b_2(M)}-1}{17^{b_2(M)}+1}\:\:.\]
\label{egyszeresen-osszefuggo}
\end{lemma}

\begin{proof}
Concerning the first assertion if $M'$ and $M''$ are as 
required then there exists an integer $k\geqq 0$ such that 
$M'\#k(S^2\times S^2)\cong M''\#k(S^2\times S^2)$ (cf. e.g. 
\cite[Theorem 9.1.12]{gom-sti}). Thus we know that we have the equality 
$\gamma (M'\#k(S^2\times S^2))=\gamma (M''\#k(S^2\times S^2))$. Then 
applying the gluing principle 
\[\frac{\gamma (M')+\gamma (k(S^2\times S^2))}{1+\gamma (M')
\gamma (k(S^2\times S^2))}=
\frac{\gamma (M'')+\gamma 
(k(S^2\times S^2))}{1+\gamma (M'')\gamma (k(S^2\times S^2))}\]
and observing that the map $x\mapsto 
\frac{x+x_0}{1+xx_0}$ is invertible we obtain that $\gamma (M')=\gamma (M'')$.

Concerning the second assertion, if $M_1$ and $M_2$ are connected, 
closed, simply connected, smooth then there exist integers $k_1,l_1\geqq 0$ and 
$k_2,l_2\geqq 0$ such that $M_1\#k_1\C P^2\#l_1\overline{\C P^2}\cong 
M_2\#k_2\C P^2\#l_2\overline{\C P^2}$ (cf. e.g. 
\cite[Theorem 9.1.14]{gom-sti}) which shows that 
$\gamma (M_1\#k_1\C P^2\#l_1\overline{\C P^2})=
\gamma (M_2\#k_2\C P^2\#l_2\overline{\C P^2})$. Now put $y:=
\gamma (\C P^2)=\gamma (\overline{\C P}^2)$. Then by the gluing principle
\[\frac{\gamma (M_1)+R_{k_1+l_1}(y)}
{1+\gamma (M_1)R_{k_1+l_1}(y)}= 
\frac{\gamma (M_2)+R_{k_2+l_2}(y)}
{1+\gamma (M_2)R_{k_2+l_2}(y)}\:\:.\]
Let $M_1:=M$ be arbitrary and 
$M_2:=S^4$ hence $\gamma (M_2)=0$. 
Then we can suppose that $k_1+l_1\leqq k_2+l_2$ therefore 
\[\frac{\gamma (M)+R_{k_1+l_1}(y)}
{1+\gamma (M)R_{k_1+l_1}(y)}=R_{k_2+l_2}(y)\]
from which again by invertability we find 
$\gamma (M)=R_{k_2+l_2-k_1-l_1}(y)$ hence setting $n:=k_2+l_2-k_1-l_1\geqq0$ 
we obtain $\gamma (M)=R_n(y)$. Moreover it is clear from the proof 
that in fact $n=b_2(M)$.

A closed formula arises if we write $R_k(y)=\frac{a_k(y)}{b_k(y)}$ and observe 
that 
\[\left(\begin{matrix}
           a_k(y)\\
           b_k(y)
 \end{matrix}\right) =
\left(\begin{matrix}
           1 & y\\
           y & 1
      \end{matrix}\right)^k
         \left(\begin{matrix}
                      0\\
                      1
                \end{matrix}\right)\:\:.\]
The eigenvalues of $\left(\begin{smallmatrix}
           1 & y\\
           y & 1
      \end{smallmatrix}\right)$ are equal to $1\pm y$ with eigenvectors 
$\left(\begin{smallmatrix}  
                      1\\
                      \pm 1
                \end{smallmatrix}\right)$ respectively. Thus inserting 
the decomposition 
$\left(\begin{smallmatrix}  
                      0\\
                      1
                \end{smallmatrix}\right)=\frac{1}{2}\left(\begin{smallmatrix}
                      1\\
                      1
                \end{smallmatrix}\right)-\frac{1}{2}\left(\begin{smallmatrix}
                      1\\
                      -1
                \end{smallmatrix}\right)$ and $k=b_2(M)$ into the above 
equation we end up with 
\[\gamma (M)=R_{b_2(M)}(y)=\frac{(1+y)^{b_2(M)}-(1-y)^{b_2(M)}}
{(1+y)^{b_2(M)}+(1-y)^{b_2(M)}}\:\:.\]

Our last task is to find the precise value of $y\in[0,1)$. Endowing 
$\C P^2$ with the orientation compatible with its complex structure and 
putting the Fubini--Study metric the group ${\rm U}(3)$ acts by 
isometries on $\C P^2$. Then the underlying orientation-preserving 
diffeomorphisms induce an action of the complexification 
${\rm U}(3)^\C={\rm GL}(3;\C)$ on $\fr$ by unitary inner 
automorphisms. Consequently, on the one hand, there exists an invariant 
subfactor $\fj (\C P^2)\subset\fr$ such that 
$\fr=\fm_3(\fj (\C P^2))$. In 
the proof of Lemma \ref{terbelilemma} we have seen that 
by exploiting the flexibility of $F_{\Sigma,\omega}$ the condition 
$A\in I(\C P^2)$ can be written as 
$F_{\Sigma, \varphi_\Sigma}(A^\divideontimes A)=
(A^\divideontimes A\varphi_\Sigma, \varphi_\Sigma)_{L^2(M,g)}=0$ where 
$\varphi_\Sigma$ is the Hodge-dual of the harmonic representative 
$\eta_\Sigma$ 
of the Poincar\'e-dual of $\Sigma\looparrowright M$ i.e., $I(\C P^2)$ 
can be defined in terms of harmonic $2$-forms only. However the 
isometries of $\C P^2$ leave the space of harmonic $2$-forms unchanged 
consequently, on the other hand, the subfactor 
$\fii (\C P^2)\subset\fr$ generated by 
$I(\C P^2)I(\C P^2)^\divideontimes\subset\fr$ as in Lemma \ref{jones} 
is invariant under ${\rm GL}(3;\C )$, too. Therefore $\fj (\C P^2)=
\fii (\C P^2)$ implying for the Jones index in question that 
\[[\fr:\fii (\C P^2)]=
[\fm_3(\fj (\C P^2)):\fj (\C P^2)
\otimes 1_{{\rm GL}(3;\C)}]=3^2=9\:\:.\] 
Thus putting together everyting we get via Lemma \ref{jones} that 
$y=R_1(y)=\gamma (\C P^2)=1-\frac{1}{9}=\frac{8}{9}$ and the 
proof is now complete.
\end{proof}

\begin{remark}\rm It follows that $\log(1-R_k(y))\approx 
k\log\frac{1-y}{1+y}$ for $k$ large enough (here ``large'' depends on 
$y$). This yields an interesting estimate for the $2^{\rm nd}$ Betti 
number of a closed simply connected manifold in terms of its 
$\gamma$-invariant. On substituting $\gamma (M)=R_{b_2(M)}(\frac{8}{9})$ 
we get 
\begin{equation} 
b_2(M)\approx\frac{\log(1-\gamma(M))}{\log\frac{1}{17}}\:\:\:\:\: 
\mbox{if $b_2(M)$ is large enough}\:\:. \label{novekedes} 
\end{equation} 
\end{remark} 
\noindent {\it Proof of Theorem \ref{nagytetel3}.} This 
theorem follows from Lemma \ref{egyszeresen-osszefuggo}.\hspace{6.5cm}$\square$ 
\newpage
%\begin{comment}
\begin{remark}\rm We have seen that 
$\fr$ is especially interesting in the four 
dimensional case since it contains curvature tensors. However actually 
this has not been used during the construction of $\gamma$ in  
the proof of Lemma \ref{terbelilemma}. Moreover, as recorded in Lemma 
\ref{egyszeresen-osszefuggo}, despite its construction depending on the 
smooth structure, $\gamma$ is in fact a topological invariant only at 
least in the simply connected closed case. Nevertheless $\gamma$ is 
computable and non-trivial. In order to appreciate $\gamma$ as it is we close 
this section by rapidly introducing another huge class of invariants whose 
members look very natural and 
meet the demands above however turn out to be trivial. 

By the aid of the canonical inclusions $C(M)\subset\ch$ and $C(M)\subset\fr
\subset\fb(\ch)$ and the Riesz representation theorem an immense class of 
functionals on $\fr$ arises by picking any $T\in\fb(\ch)$ and continuously 
extending the map $A\mapsto (\hat{T},\widehat{A^\divideontimes})=\tau(TA)$ from 
$C(M)$ to $\fr$ giving rise to a continuous $\C$-linear map 
$F_T:\fr\rightarrow\C$ of the form 
\[F_T(A):=\tau (TA)\]
which is also positive if $\tau(T)>0$. 
Interesting examples emerge if instead of a plain manifold $M$ as in 
Lemma \ref{terbelilemma} we start now with an oriented Riemannian one $(M,g)$. 
For instance take the Hodge operator $*$ induced by the metric and 
orientation and put $T:=\frac{1}{2}(1+*)$ hence $TA=A^+$ is the 
``self-dual part'' of $A$. Or assume that the curvature is bounded 
hence $R_g\in\fr$ and satisfies $\tau (R_g)\not=0$ i.e. the metric itself 
is sufficiently generic\footnote{Note 
that whenever $M$ is compact it can accommodate a metric $g$ for which either 
$\tau (R_g)\not=0$ or $\tau (R_g)=0$ 
i.e. both cases can occur. This is obvious for flat manifolds admitting a 
metric with $R_g=0$ hence $\tau(R_g)=0$. 
%(The list of compact flat $4$-manifolds is available in \cite{hil}.) 
But note that (\ref{nyomok}) says that $\tau (R_g)$ is proportional to the 
total scalar curvature hence $\tau (R_g)=0$ for all Ricci-flat manifolds too. 
But even more if $M$ is any compact oriented manifold then take a smooth 
function $f:M\rightarrow\R$ which is (i) strictly
negative in one point and (ii) satisfies $\int_Mf\mu_h=0$ with respect
to some auxiliary metric $h$. Then (i) implies by \cite[Theorem 4.35]{bes}
the existence of a Riemannian metric $g$ having $f$ as its scalar curvature
hence via (ii) and (\ref{nyomok}) satisfying $\tau (R_g)=0$.} and put 
$T:=\pm R_g$. Let $\{0\}\subseteqq I(M)\subsetneqq\fr$ denote the usual 
multiplicative left-ideal generated by this functional i.e. $A\in I(M)$ if 
and only if $F_T(A^\divideontimes A)=0$. Now if $T'$ is another element 
satisfying $\tau (T')\not=0$ then by assumption both 
$0\not=\hat{T},\hat{T'}\in\ch$ hence there 
exists an invertible element $S\in\fb(\ch)^\times$ such that $T'=ST$; 
consequently 
\[0\leqq\vert F_{T'}(A^\divideontimes A)\vert^2=\vert\tau 
(STA^\divideontimes A)\vert^2=\vert
\tau (TA^\divideontimes AS)\vert^2
=\vert F_T(A^\divideontimes AS)\vert^2\leqq 
F_T(A^\divideontimes A)F_T((AS)^\divideontimes AS)\]
and likewise $0\leqq \vert F_T(A^\divideontimes A)\vert^2\leqq 
F_{T'}(A^\divideontimes A)F_{T'}((AS^{-1})^\divideontimes AS^{-1})$. This 
demonstrates the independence of $I(M)$. Take again the self-adjoint subspace 
$I(M)I(M)^\divideontimes$ and let $\{0\}\subseteqq\ci(M)\subsetneqq\ch$ be 
the closure of $C(M)\cap I(M)I(M)^\divideontimes$ and consider its orthogonal 
complementum $\{0\}\subsetneqq\ci(M)^\perp\subseteqq\ch$ and the corresponding 
projection $P_M:\ch\rightarrow\ci (M)^\perp$. Also take the 
subfactor $\C\cong Z(\fr)\subseteqq\fii (M)\subsetneqq\fr$ provided by the 
weak closure of $C(M)\cap I(M)I(M)^\divideontimes$. Very similarly to the 
construction in the proof of Lemma \ref{terbelilemma} we observe that 
$\ci(M)^\perp$ is a left $\fr$-module; hence define a new smooth $4$-manifold 
invariant by its coupling constant more precisely 
\[\delta(M):=\tau (P_M)\:\:.\] 
The immediate observation is that $\delta(M)\in(0,1]$. However 
the analogue of Lemma \ref{jones} is not valid because $\ci(M)$ is 
not necessarily the standard $\fii (M)$-module anymore as the case of 
$\ci(M)=\{0\}$ and the corresponding $\fii (M)=Z(\fr)\cong\C$ shows. The 
computationally useful Lemma \ref{gammaformulak} also no longer holds 
(except the invariance under reversing orientation). In fact since we can 
suppose that $T$ is self-adjoint and positive definite therefore 
$T=S^2$ with a same type hence invertible operator 
$S\in\fb(\ch)$; thus using the norm $\Vert\:\cdot\:\Vert$ on $\ch$ we find 
$0=F_T(A^\divideontimes A)=\Vert \widehat{(AS)^\divideontimes}\Vert^2$ implying 
$AS=0$ i.e. $A=0$. Thus $I(M)=0$ yielding $\ci(M)^\perp=\ch$. Thus the 
fact is that always $\delta(M)=1$. 
\end{remark}
%\end{comment}

%%%%%%%%%%%%%%%%%%%%%%%%%%%%%%%%%%%%

\section{Physical interpretation}
\label{three}

%%%%%%%%%%%%%%%%%%%%%%%%%%%%%%%%%%%%

In this section we cannot resist temptation and introduce the basics of 
a physical interpretation of the material collected so far. Namely, 
upon reversing the approach of Section \ref{two}, we shall replace the
immense class of classical space-times of general relativity with a single
universal ``quantum space-time'' allowing us to lay down the foundations
of a manifestly four dimensional, covariant, non-perturbative and
genuinely quantum theory of gravity. This construction is natural, simple and  
self-contained. More precisely here in Section \ref{three} not 
{\it one particular $4$-manifold}---physically regarded as a particular 
classical space-time---but {\it the unique hyperfinte ${\rm II}_1$ factor 
von Neumann algebra}---physically viewed as the universal quantum 
space-time---is declared to be the primarily given object. Let us see 
how it works.

{\it Observables, fields, states and the gauge group}. Let $\ch$ be an
abstractly given infinite dimensional complex separable Hilbert space and
$\fr\subset\fb (\ch)$ be a type ${\rm II}_1$ hyperfinite factor hence tracial 
von Neumann algebra acting on $\ch$ by the standard representation.
We call $\fr$ the {\it algebra of (bounded) observables},
its tangent space $T_1\fr\supset\fr$ consisting of the Fr\'echet derivatives of
$1$-parameter families of observables at the unit $1\in\fr$ the {\it
algebra of fields}, while $\ch$ the {\it state space} in this quantum
theory. The subgroup ${\rm U}(\ch)\cap\fr$ of the unitary group
of $\ch$ operating as inner automorphisms on $\fr$ is
the {\it gauge group}. Note that the gauge group acts
on both $\fr$ and $\ch$ but in a different way.

What kind of quantum theory is the one in which 
$\fr$ plays the role of the algebra of {\it physical observables}? We 
have seen in Section \ref{two} that $\fr$, when regarded to operate on a 
differentiable manifold, contains all bounded (complexified)
algebraic curvature tensors along it whenever the real
dimension of this manifold is precisely {\it four}. Moreover $\fr$ 
is generated by local operators more precisely by trace class algebraic 
operators $R$ acting like {\it multiplication} operators and by trace class 
operators $L_X$ acting like {\it first order differential} 
operators. Picking canonically conjugate pairs i.e. necessarily unbounded 
self-adjoint operators $R,L_X\in T_1\fr$ satisfying $[R,L_X]=1$ 
(assuming $\hbar=1$)
%\footnote{Such pairs are provided by picking a real 
%function $f$ and a real vector field $X$ satisfying $Xf=1$ and taking 
%$R:=M_f$ and $L_X$. Note that the equation $Xf=1$ has no bounded solution.} 
shows that $T_1\fr$ contains a CCR algebra hence 
revealing the {\it bosonic} character of $\fr$. Recall that in classical 
general relativity {\it local} gravitational phenomena are caused by the 
curvature of space-time. Hence interpreting $\fr$ physically as an operator 
algebra of local observables the corresponding quantum theory is expected to 
be a {\it four dimensional quantum theory of pure gravity} (hence a bosonic 
theory). In this way we fulfill the {\it Heisenberg dictum} that a 
quantum theory should completely and unambigously be formulated and interpreted 
in terms of its local physical observables (and not the other way round). 
In modern understanding by a {\it physical theory} one means a two-level 
description of a bundle of phenomenologically interrelated 
natural phenomena: the theory possesses a {\it syntax} provided by its 
mathematical core structure and a {\it semantics} which is the meaning 
i.e. interpretation of the bare mathematical model in terms of physical 
concepts. In this context our quantum 
theory is not merely a mathematical theory anymore but a physical 
theory. This is because the bare mathematical structure $\fr$ (together 
with a representation $\pi$ on $\ch$) is dressed up i.e., interpreted by 
assigning a physical (in fact, gravitational) meaning to the experiments 
consistently performable by the aid of this structure (i.e., the usual 
quantum measurements of operators $A\in\fr$ in pure states $v\in\ch$ or 
in more general ones, see below). In our opinion it is of particular 
interest that the geometrical dimension---equal to four---is fixed at 
the semantical level only and it matches the known phenomenological 
dimension of space-time. This is in sharp contrast to e.g. string theory 
where the geometrical dimension of the theory is fixed already at its 
syntactical level i.e., by its mathematical structure (namely, 
demaninding the theory to be free of conformal anomaly) and it turns out 
to be much higher than the phenomenological dimension of space-time. 
As a further clarification it is emphasized that the aim here is {\it not} to 
quantize general relativity in some way but rather conversely: being 
a successful theory at the classical level, general relativity 
should be derived from this abstractly given quantum 
theory by taking an appropriate ``classical limit'' as naturally as possible. 
Unfortunately, this program yet cannot be carried out completely here. 

{\it Observables as the universal space of all space-times}. Taking 
into account the embeddings (\ref{beagyazas}) and the universality of $\fr$ 
the {\it unique} algebra of observables $\fr$ can be considered 
as the collection of {\it all} classical space-times and we can interpret the 
appearance of the gauge group as the manifestation of the diffeomorphism 
gauge symmetry of classical general relativity in this quantum theory. 
Indeed, orientation-preserving diffeomorphisms of $M$ 
interchange its points as well as act on the corresponding operators 
in $\fr$ (projections) by unitary inner automorphisms i.e. ${\rm Diff}^+(M)$ 
embeds into ${\rm Aut}\:\fr$. Reformulating
this in a more geometric language we can say that classical space-times
appear as special orbits within $\fr$ of its gauge group. The operators in 
$\fr$ representing geometric points via (\ref{beagyazas}) are 
special operators namely projections. Consequently 
the full non-commutative algebra $\fr$ is not exhausted by operators
representing points of space-time; it certainly contains much more
operators---e.g. various projections which are not of geometric
origin---therefore this ``universal quantum space-time'' is more than a
bunch of all classical space-times. As a comparison, in 
{\it algebraic quantum field theory} \cite{haa}
one starts with a particular smooth $4$-manifold $M$ and considers an
assignment $U\mapsto\fr (U)$ describing local algebras of observables
along all open subsets $\emptyset\subseteqq U\subseteqq M$. However in our
case, quite conversely, space-times are secondary structures only and
all of them are injected into the unique observable algebra $\fr$ which
is considered to be primary.

{\it More examples of observables, fields and states}. Let us take a 
closer look of the elements of $\fr$ and of $T_1\fr\supset\fr$. Taking 
into account the items above concerning the interpretation of the 
mathematical results of Section \ref{two} we have agreed to identify 
some elements of $\fr$ up to finite accuracy with four dimensional 
bounded algebraic (i.e., formal) curvature tensors of {\it all possible} 
smooth $4$-manifolds. In fact one can more accurately reformulate some 
basic concepts of classical general relativity within this quantum 
(field-)theoretic framework. The only thing we have to do is to refine 
or improve the constructions in Section \ref{two} by taking into account 
not only the non-degeneracy but the indefiniteness of the pairing 
(\ref{integralas}) too.

Remember that given a connected oriented Riemannian $4$-manifold $(M,g)$
one can introduce the Hodge operator $*$ which satisfies $*^2=
{\rm Id}_{\Omega^2_c(M)}$ hence it induces a splitting
\[\Omega^2_c(M)=\Omega^+_c(M)\oplus\Omega^-_c(M)\]
into (anti)self-dual or $\pm1$-eigenspaces. Thus the Hodge star as an
operator on $\Omega^2_c(M)$ with respect to this splitting
has the form $*=\left(\begin{smallmatrix}
                    1&0\\
                    0&-1
         \end{smallmatrix}\right)$ hence is symmetric; likewise the
curvature tensor $R_g$ of $(M,g)$ looks like (\ref{gorbulet})
i.e., is also a pointwise symmetric operator. We summarize these facts by
writing  that
$*,R_g\in C^\infty (M;S^2\wedge^2T^*M)\subset C^\infty(M;\ed(\wedge^2T^*M))$.
In addition $R_g$ obeys the algebraic Bianchi identity $b(R_g)=0$. The
vacuum Einstein equation with cosmological
constant $\Lambda\in\R$ reads as ${\rm Ric}=\Lambda g$ therefore it is
equivalent to the vanishing of the traceless Ricci part ${\rm Ric}_0$ of
$R_g$ in (\ref{gorbulet}). From these it readily follows, as noticed in
\cite{sin-tho}, that $(M,g)$ is Einstein i.e., satisfies the vacuum
Einstein equation with cosmological constant if and only if
$*R_g=R_g*$ or equivalently $*R_g*=R_g$.

Now if $R_g$ is a bounded operator then we have seen in Section \ref{two}
that its complex-linear extension gives rise to an element
$R_g\in \fr$ which is moreover self-adjoint i.e.,
$R_g^\divideontimes =R_g$ because $R_g$ is symmetric.
Therefore its trace is always real. Likewise for $*$ satisfying $\tau(*)=0$.
The algebraic Bianchi identity has the form $b(R_g)=0$ where
$b$ is a certain fiberwise averaging operator on symmetric bundle morphisms
satisfying $b^2=b$ and $b^*=b$ with respect to the metric. This gives rise
to a $g$-orthogonal decomposition
$S^2\wedge^2T^*_xM\otimes\C={\rm ker}\:b_x\oplus
{\rm im}\:b_x$ at every point $x\in M$. A further subtlety of {\it
four dimensions} is that
${\rm im}\:b_x=\C*{\rm Id}_{\wedge^2T_x^*M\otimes\C}\subset
S^2\wedge^2T_x^*M\otimes\C$ holds \cite{sin-tho}.
Consequently $b(R_g)=0$ is equivalent to saying that $g(R_g,*)=0$ i.e.,
within $S^2\wedge^2T^*M\otimes\C$ the operator $R_g$ is $g$-orthogonal
to $*$ at every $x\in M$. By exploiting
that $*$ is a pointwise symmetric operator we can use the
pointwise equality $g(R_g,*)={\rm tr}(R_g*)$ and then (\ref{nyomok}) to obtain
\[\tau(R_g*)=\frac{1}{6}\int\limits_M{\rm tr}(R_g*)\mu_g=0\]
which geometrically means that $(\hat{R}_g,\hat{*})=0$ i.e., $\hat{R}_g$ is
perpendicular to $\hat{*}$ within $\ch$. Hence the Bianchi identity.
Concerning the Einstein condition if $R_g$ comes from an Einstein metric
with cosmological constant
$\Lambda\in\R$ then we already know that $*R_g*=R_g$ which means that
${\rm Ric}_0=0$ in (\ref{gorbulet}). Moreover taking into
account that ${\rm Weyl}^\pm\in C^\infty(M;S^2\wedge^2T^*M\otimes\C)$
in (\ref{gorbulet}) are traceless their pointwise scalar products with
the identity of $C^\infty(M;S^2\wedge^2T^*M\otimes\C)$ look like
$g({\rm Weyl}^\pm\:,1)={\rm tr}({\rm Weyl}^\pm)=0$. Using
(\ref{nyomok}) this shows again that the corresponding
vectors $\widehat{\rm Weyl}^\pm\in\ch$ are perpendicular to
$\hat{1}\in\ch$ i.e., $(\widehat{\rm Weyl}^\pm\:,\:\hat{1})=0$.
Moreover ${\rm Scal}=4\Lambda$ consequently via (\ref{gorbulet}) we get
\begin{equation}
\tau (R_g)=(\hat{R}_g\:,\:\hat{1})=\frac{1}{12}(\widehat{\rm Scal},\hat{1})
=\frac{4\Lambda}{12}(\hat{1},\hat{1})=\frac{\Lambda}{3}\:\:.
\label{energia}
\end{equation}
Finally note that in the previous analysis the signature of the
metric plays no significant role because of an involved complexification
(we have used Riemannian signature just for convenience).

All of these serve as a motivation for the following operator-algebraic
reformulation or generalization of the classical vacuum Einstein equation
with cosmological constant:

\begin{definition}
Let $M$ be a connected oriented smooth $4$-manifold and $\fr$
its ${\rm II}_1$-type hyperfinite factor von Neumann algebra as before.

(i) A {\rm refinement of $\fr$} is a pair $(\fr,*)$ where
$1\not=*\in\fr$ and satisfies $*^2=1$;

(ii) An operator $Q\in\fr$ {\rm solves the quantum vacuum Einstein
equation} with respect to $(\fr,*)$ if
\[\left\{\begin{array}{lll}
Q^\divideontimes &=Q & \mbox{(self-adjointness)}\\
\tau (Q*)&=0 & \mbox{(algebraic Bianchi identity)}\\
*Q*& =Q & \mbox{(Einstein condition);}
\end{array}\right.\]

(iii) The trace $\tau(Q)=:\frac{\Lambda}{3}\in\R$ is
called the corresponding {\rm quantum cosmological constant};

(iv) $Q\in\fr$ as above is called a {\rm  vacuum state} 
if it is positive semi-definite too (hence its corresponding quantum 
cosmological constant is non-negative).   
\label{kvantumeinstein}
\end{definition}

\noindent Note that the curvature tensor $R_g$ of an Einstein manifold
$(M,g)$ if bounded as an operator always solves the quantum vacuum Einstein 
equation with respect to the canonical refinement provided by metric
(anti)self-duality. However in sharp contrast to the classical Einstein
equation whose solution is a metric therefore is non-linear, its quantum
generalization is linear hence easily solvable. Of course this is beacuse
in the generalization we do not demand the solution (which can be any
non-local operator) to originate from a metric. Indeed, given $B\in\fr$
then $S=\frac{1}{2}(B+B^\divideontimes)$ is self-adjoint and then
picking an arbitrary refinement $(\fr,*)$ and taking into account
that $*$ is always self-adjoint, the averaged operator 
$Q:=\frac{1}{2}\big(S+*S*\big)-\tau(S*)*\in\fr$ is automatically a solution 
of the quantum vacuum Einstein equation; moreover using $\tau(*)=0$ its trace 
$\frac{\Lambda}{3}$ is equal to $\tau (B)$ hence is independent of the 
particular refinement. Observe that this linearity permits a BRST-like 
reformulation of Definition \ref{kvantumeinstein} too i.e. introducing a linear 
operator $\Delta:\fr\rightarrow\fr$ such that ${\rm ker}\Delta$ is precisely 
equal to the solution space and $\Delta^2=0$.

It is important to note that, contrary to smooth solutions, many 
singular solutions of classical general relativity theory cannot be 
interpreted as observables because their curvatures 
lack being bounded operators hence do not belong to $\fr$. As a result,
we expect that the classical Schwarzschild or Kerr black
hole solutions and more generally gravitatioal fields of isolated bodies 
(cf. \cite{ein,ein-pau}) give rise not to {\it observables} in $\fr$ but
rather {\it fields} in $T_1\fr\supset\fr$. 

{\it Questions and answers.} First let us clarify what the answers in 
this proposed quantum theory are because this is easier. Staying safely 
within the orthodox framework i.e., the {\it Copenhagen interpretation} and the
standard mathematical formulation of quantum theory (but relaxing this
latter somewhat), given an observable represented by $A\in\fr$ and a
general (i.e., not necessarily pure) state also represented by an element
$B\in\fr$ in the observable algebra (regarded as a ``density matrix''
operator over the state space $\ch$) we declare that an {\it answer} is 
formally like
\[\mbox{{\it The expectation value of the observable quantity $A$ in the
state $B$ is $\tau (AB)\in\C$}}\]
where $\tau :\fr\rightarrow\C$ is the
unique finite trace on the hyperfinite ${\rm II}_1$ factor von Neumann
algebra $\fr$. Note that in order not to be short sighted, at this level of
generality we require neither the observable $A\in\fr$ to be self-adjoint 
nor the state $B\in\fr$ to
be positive semi-definite self-adjoint and normalized (however these can be 
imposed if they turn out to be necessary, cf. e.g. the characterization 
of a vacuum state in Definition \ref{kvantumeinstein} above) hence our 
answers can be complex 
numbers in general. Nevertheless $\tau (AB)$ is well-posed i.e., is finite and
invariant under the gauge group of this theory namely the unitary
automorphisms of $\fr$ thus it is indeed an ``answer''---at least syntactically.

Now we come to the most difficult problem namely what are the meaningful 
questions here? This problem is already fully at
the foggy semantical level. The orthodox approach says that a {\it question} 
should formally be like
\begin{eqnarray}
&&\mbox{{\it From the collection ${\rm Spec}A\subset\C$ of
``all possible values the pointer of the experimental}}\nonumber\\
&&\mbox{{\it instrument designed to measure $A$ in
the laboratory can assume'', which does occur in $B$?}}\nonumber
\end{eqnarray}
and the answer (as defined before) to this question is obtained through 
a {\it measurement}. Thus let us make a short interlude concerning the 
measurement. After performing the physical experiment designed to answer 
the question above should we expect that the state $B\in\fr$ will 
necessarily ``collapse'' to an eigenstate $B_\lambda\in\fr$ of $A$? In 
our opinion {\it no} and this is an essential difference between gravity 
and quantum mechanics. Namely, in quantum mechanics an ideal observer 
compared to the physical object to be observed is {\it infinitely large} 
hence the immense physical interaction accompanying the measurement 
procedure drastically disturbs the entity leading to the collapse of its 
state. However, in sharp contrast to this, in gravity an ideal observer 
is {\it infinitely small} hence it is reasonable to expect that 
measurements might {\it not} alter gravitational states. This is in 
accordance with our old experience concerning measurements in astronomy.

Concerning the problem of its {\it meaning}, since $A,B\in\fr$ have 
something to do with the curvature of local portions of space-time, it 
is not easy to assign a straightforward meaning to the above question. 
Therefore instead of offering a general solution to this problem at this 
initial state of the art, let us rather consider some special cases. For 
example if $(M,g)$ is a classical non-singular space-time and $(N, h)$ 
is another one, then their {\it classical geometrical} habitants may 
find that $0\not=\tau (R_gR_h)\in\C$ in general. Consequently ``physical 
contacts'' between different classical geometries can already occur 
(whatever it means). It follows from the construction of the trace in 
(\ref{nyomok}) that $\tau (R_gR_h)$ arises by integrating the local 
trace function ${\rm tr}(R_gR_h)$ along $X=M\cap N$. Note that making 
use of the embeddings of $M,N$ into $\fr$ given by (\ref{beagyazas}) 
taking intersection (maybe empty) is meaningful. Thus given a nearly 
flat space-time $(M,g)$ ``we live in'' i.e., an observer satisfying 
$\vert R_g(x)\vert_{g}\approx 0$ for all $x\in M$ and a different 
geometry $(N,h)$ ''we observe'' i.e., a state $R_h$ then we expect 
$\vert\tau (R_gR_h)\vert\approx 0$ in accord with our physical intuition 
that frequent encounters with different geometries in quantum gravity 
should occur rather in the strong gravity regime of space-times (hence 
the reason we do not experience such strange things).

As another example for measurement, consider the energy. In this universal 
quantum theory the only distinguished non-trivial self-adjoint gauge invariant 
operator is the identity $1\in\fr$. Therefore the only natural candidate for 
playing the role of a {\it Hamiltonian} responsible for dynamics in this theory 
is $1\in\fr$ (in natural units $c=\hbar =G=1$). Therefore this dynamics is 
trivial. Nevertheless, quite interestingly, it coincides with the {\it 
modular dynamics} introduced by Connes and Rovelli \cite{con-rov} because
it is associated with the {\it tracial} state $\tau$ on $\fr$ having the
identity as its modular operator. Therefore the odd-looking 
equality (\ref{energia}) can be interpreted by saying that the 
expectation value (in a sequence of measurements) of the energy 
in the state represented by a vacuum curvature 
tensor $R_g$ is the number $\tau (R_g)$ and this energy is equal to the 
cosmological constant constituent $\frac{\Lambda}{3}$ of the corresponding 
vacuum space-time $(M,g)$. Note that $R_g$ is at least self-adjoint 
(but perhaps not positive definite neither normalized) hence the energy is a 
real number but {\it a priori} can assume negative values too, cf. item 
(iv) of Definition \ref{kvantumeinstein}. 

{\it An application to the cosmological constant problem.} The last worry 
concerning the non-positivity of the energy strongly motivates to test this 
formalism on the so-called cosmological constant problem. By this we mean the 
complex problematics introduced into current 
standard cosmology (based on the cosmological principle and the 
corresponding Friedmann--Lema\^{\i}tre--Robertson--Walker or FLRW model) 
and standard particle physics (based on the Standard Model) by the 
experimental verification \cite{rie-etal} of the existence of a 
strictly {\it positive} but {\it small} cosmological constant in 1998. Recall 
that in cosmology the cosmological constant is defined as 
$\Lambda=3\left(\frac{H_0}{c}\right)^2\Omega_\Lambda$ where $H_0$ is the
Hubble ``constant'' (it actually varies with time) whose dimenion is 
${\rm s}^{-1}$, $c$ is the speed of light having dimension ${\rm ms}^{-1}$ 
and the dimensionless number $\Omega_\Lambda\approx 0.69$ is the 
ratio of the dark energy density and the critical material energy density 
($\Omega_\Lambda$ also changes in time). Thus the dimension of the physical 
cosmological constant is ${\rm m}^{-2}$. Based on astronomical 
measurements $\Lambda\approx 2.89\times 10^{-122}\:\ell_{\rm Planck}^{-2}$ 
where $\ell_{\rm Planck}\approx 1.62\times 10^{-35}$ m. In order to gain 
compatibility with the mathematical expressions obtained in Section \ref{two}, 
from now on we shall use the dimensionless number 
\[\frac{\Lambda}{3}\ell_{\rm Planck}^2\approx 0.96\times 10^{-122}\] 
expressing the magnitude of the cosmological constant in natural Planck units. 
We claim that this sort of (i.e., a strictly positive but small) 
cosmological constant naturally appears if classical general relativity is 
replaced with its formal quantum generalization as introduced above. 
The cosmological constant problem has been investigated from a 
closely related angle by other authors too, cf. \cite{ass-kro}.

Before seeing this however let us briefly and roughly clarify what do we 
mean by this ``replacement'' from a physical viewpoint. We also want 
to see how a cosmological context has popped up so suddenly. The modern 
physical and mathematical basis for thinking about space, time, 
gravity and the structure and history of the Universe (cosmology) has 
been provided by classical general relativity in 
the form of a pseud-Riemannian $4$-manifold $(M,g)$ such that 
$g$ is subject to Einstein's field equation. We have seen in 
Section \ref{two} especially in Lemma \ref{terbelilemma} 
that out of this input data (in fact from $M$ alone) an operator 
algebra $\fr$ and a Hilbert space $\ci(M)^\perp$ carrying a 
representation $\rho_M$ of $\fr$ can be constructed. 
Moreover the representation $\rho_M$ on $\ci(M)^\perp$ is 
the restriction of the unique standard representation $\pi$ of $\fr$ 
on a Hilbert space $\ch$ containing $\ci(M)^\perp$. The latter objects 
namely $\fr,\ch,\pi$ are all {\it unique} up to corresponding 
isomorphisms. The two structures (namely the ``classical'' $(M,g)$ on the 
one side 
and the ``quantum'' $(\fr,\ch,\pi)$ on the other) are naturally connected 
mathematically by the observation made in Section \ref{two} that $\fr$ 
contains bounded curvature tensors and by Definition \ref{kvantumeinstein} 
offering a re-interpretation of the vacuum Einstein equation in an 
operator-theoretic language. Remember 
the plain technicality that a curvature tensor must be {\it bounded} in 
order to belong to the operator algebra. However it has been known for a long 
time that in general relativity the gravitational field of an isolated 
massive configuration cannot be regular everywhere \cite{ein, ein-pau}. Thus we 
are forced to move towards a global non-singular cosmological context when 
trying to work out a physical connection between the two structures. Therefore 
hereby we make a {\it working hypothesis}: the unique abstract triple 
$(\fr,\ch,\pi)$ as a quantum 
theory describes the Universe from the Big Bang till the Planck time 
$t_{\rm Planck}\approx 5.39\times 10^{-44}$ s while a highly 
non-unique pair $(M,g)$ as a space-time in classical general relativity 
describes its evolution from $t_{\rm Planck}$ onwards. 
A usual choice for $(M,g)$ is the FLRW solution with or without 
cosmological constant. The ``moment'' $t_{\rm Planck}$ would therefore 
formally or symbolically label the emergence or onset of a space-time 
structure in the course of the history of the Universe\footnote{We are
aware of, but cannot do better, how nonsense it is, strictly speaking, to
talk about the emergence of space and time at the moment $t_{\rm Planck}$;
this assertion requires the contradictory pre-supposition of time before
$t_{\rm Planck}$.} establishing a physical correspondence 
between the two aforementioned mathematical structures. We symbolically 
write this correspondence as 
\[\begin{matrix}
(\fr,\ch,\pi) & \Longrightarrow & \big(Q_M\in \fr, \ci(M)^\perp\subset\ch,
\rho_M=\pi\vert_{\ci(M)^\perp}\big) & \Longleftrightarrow & M\\ 
& \Longrightarrow & \big(Q_M,R_g\in\fr,\ci(M)^\perp\subset\ch,\rho_M
=\pi\vert_{\ci(M)^\perp}\big) &\Longleftrightarrow & (M,g)
\end{matrix}\] 
and having a sort of spontaneous symmetry breaking or phase transition in mind 
occuring at $t_{\rm Planck}$. Actually we broke this process into two parts: 
``first'' the emergence of the plain manifold structure $M$ 
provided by a choice of an (in our convention dimensionless) operator 
$Q_M\in\fr$ satisfying the quantum vacuum Einstein equation as in 
Definition \ref{kvantumeinstein} and 
``then'' the emergence of the geometry $g$ hence the causal structure 
over $M$ too provided by a further (in our convention dimensionless) 
curvature operator $R_g=Q_M+8\pi T\in\fr$ satisfying the classical 
Einstein equation with already fixed cosmological constant 
$\frac{\Lambda}{3}\ell^2_{\rm Planck}=\tau(Q_M)$ and 
some matter content described by a further (in our convention dimensionless) 
stress-energy tensor $T$. This passage from the unique 
quantum regime $(\fr,\ch,\pi)$ to a particular classical regime $(M,g)$ 
could probably be more rigorously captured in the framework of algebraic 
quantum field theory \cite{haa} as switching from the unique representation 
$\pi$ to a different particular representation $\rho_M$ of $\fr$. 
In this framework one can also formally label the transition with 
$T_{\rm Planck}\approx1.42\times 10^{32}$ K; this high temperature is 
reasonable if we make a 
further technical observation  that $\pi$ is induced by the unique tracial 
state $\tau$ on $\fr$ hence one can imagine that $(\fr,\ch,\pi)$ describes a 
quantum statistical mechanical system being in an infinitely high temperature 
Kubo--Martin--Schwinger or KMS equilibrium state $\tau$ (hence 
having trivial modular dynamics in the sense of \cite{con-rov}).

After this admittedly yet incomplete trial to set up some physical picture 
we can embark upon the treatment of the cosmological constant problem in this 
framework. Let $M$ be a connected oriented smooth $4$-manifold and consider 
its associated projector $P_M\in\fr$ from Lemma 
\ref{terbelilemma}. Recall that an operator like $P_M$ has been used to 
encode in the simplest way a particular connected oriented smooth 
$4$-manifold $M$ within its operator algebra $\fr$ which is however 
universal. Thus $P_M$ is a distinguished operator in this sense. But even 
more, it is self-adjoint hence taking an arbitrary refinement $(\fr,*)$ 
its average 
\[Q_M:=\frac{1}{2}\big(1-P_M+*(1-P_M)*\big)-\tau((1-P_M)*)*
=1-\frac{1}{2}(P_M+*P_M*)+\tau(P_M*)*\] 
is an operator which solves the quantum vacuum Einstein equation too in 
the sense of Definition \ref{kvantumeinstein} with quantum cosmological 
constant 
$\frac{\Lambda}{3}\ell_{\rm Planck}^2=\tau (Q_M)$. The particular 
choice for $Q_M$ is verified by Lemma \ref{terbelilemma} making it sure that 
$\tau (Q_M)=\tau(1-P_M)=1-\gamma (M)$ thus 
\[\frac{\Lambda}{3}\ell_{\rm Planck}^2=1-\gamma (M)\] in terms 
of the dimensionless $\gamma$-invariant of $M$. The further 
restriction imposed upon the spectrum of $\gamma (M)$ in Lemma \ref{jones} 
guarantees that actually 
\[\frac{\Lambda}{3}\ell_{\rm Planck}^2\in\left(0,\frac{1}{4}\right]\bigcup
\left\{\frac{1}{4}\cos^{-2}\left(\frac{\pi}{n}\right)\:\Bigg\vert\: 
n=\dots5,4,3\right\}\subset (0,1]\] 
thus it is permitted to be arbitrary close but never equal to zero. That is, 
whatever $M$ is, in our model {\it the (dimensionless) cosmological constant 
is always a small positive number} (thus, using again the terminology of 
Definition \ref{kvantumeinstein}, in fact $Q_M$ is a vacuum state). This is 
consistent with the aforementioned emprical value 
$\approx0.96\times 10^{-122}$ of the cosmological constant. 

To be more specific we can impose some conditions on $M$ still 
compatible with cosmological experience. We can suppose that $M$ has the 
structure $X\setminus\{{\rm point}\}$ where $X$ is a connected 
simply connected closed smooth $4$-manifold; this is the case for 
the FLRW model in the relevant cases $k=0,-1$. Moreover we can suppose that 
the $2^{\rm nd}$ Betti number of $X$ is non-trivial: 
$b_2(X)>0$ implying $b_2(M)>0$ too. This assumption does not hold in 
FLRW cosmology but it can be physically 
interpreted as the necessary topological condition for $M$ i.e., the 
space-time to contain black holes at $t_{\rm Planck}$. 
Indeed, taking e.g. a survey on known solutions
\cite{ste-kra-mac-hoe-her}, we can make an interesting observation that 
apparently all explicitly known $4$ dimensional black hole solutions in 
vacuum general relativity have the property that
because of some general reason their black hole content is even topologically
recognizable as a two dimensional ``hole'' in space-time. This means that 
their instantaneous event horizons represented by immersed connected 
oriented surfaces in $M$ give rise to non-trivial elements in 
$H_2(M;\Z)^{\rm free}\cong\Z^{b_2(M)}$. One can indeed prove this at least 
for stationary black holes \cite{ete1}. Using this picture the number 
$b_2(M)$ can therefore be physically 
interpreted as the number of primordial black holes in the space-time at 
its onset. The assumption $0<b_2(M)$ implies that strictly 
$0<\frac{\Lambda}{3}\ell_{\rm Planck}^2\lvertneqq 1$ such that letting 
$b_2(M)\rightarrow+\infty$ we find 
$0\leftarrow\frac{\Lambda}{3}\ell_{\rm Planck}^2$. Consequently, the 
observed magnitude of the 
cosmological constant allows us to estimate in this model the number 
$N\approx b_2(M)$ of primordial black holes just around the Planck 
era. Plugging the empirical value of the cosmological constant into 
(\ref{novekedes}) we find
\[N\approx\frac{\log(\frac{\Lambda}{3}\ell_{\rm Planck}^2)}
{\log\frac{1}{17}}\sim 10^2\]
yielding a number which is negligable. This qualitative 
result obtained by topological means is compatible with the agreement 
in the physicist literature that primordial black holes are essentially 
absent from the very early Universe, cf. e.g. \cite{car-kuh,gre} 
(and the hundreds of references therein). 
Note that the usual considerations leading to this conclusion 
are based on the application of the Press--Schechter 
mechanism \cite{pre-sch} for primordial black hole 
formation; hence our topological result is an independent 
confirmation of these considerations. 
\vspace{0.1in}

\noindent{\bf Acknowledgements}. The author is grateful to T. 
Asselmeyer-Maluga, B. Gyenti and I. Nikolaev for the 
stimulating discussions. There are no conflicts of interest to declare
that are relevant to the content of this article.
The work meets all ethical standards applicable here.
All the not-referenced results in this work are fully the author's own
contribution. No funds, grants, or other financial supports were received. 
Data sharing is not applicable to this article as no datasets were generated or
analysed during the underlying study.

\end{document}